\documentclass[apj]{emulateapj}
\shortauthors{Cui \& Bai 2019}
\usepackage{amsmath, physics, hyperref, float}
\hypersetup{colorlinks=true, linkcolor=blue, citecolor = blue, urlcolor = blue}
\newcommand{\pa}{\partial}

\begin{document}
\title{Global Simulations of the Vertical Shear Instability with Non-ideal Magnetohydrodynamic Effects}
\author{Can Cui$^1$ and Xue-Ning Bai$^{2,3}$}
\affil{$^1$Shanghai Astronomical Observatory, Chinese Academy of Sciences, Shanghai 200030, China \\ 
$^2$Institute for Advanced Study, Tsinghua University, Beijing 100084, China \\
$^3$Department of Astronomy, Tsinghua University, Beijing 100084, China \\
\href{mailto:ccui@shao.ac.cn}{ccui@shao.ac.cn}; \href{mailto:xbai@tsinghua.edu.cn}{xbai@tsinghua.edu.cn}}

\begin{abstract}

The mechanisms of angular momentum transport and the level of turbulence in protoplanetary disks (PPDs) are crucial for understanding many aspects of planet formation. In recent years, it has been realized that the magneto-rotational instability (MRI) tends to be suppressed in PPDs due to non-ideal magnetohydrodynamic (MHD) effects, and the disk is primarily laminar with accretion driven by magnetized disk winds. In parallel, several hydrodynamic mechanisms have been identified that likely also generate vigorous turbulence and drive disk accretion. In this work, we study the interplay between MHD winds in PPDs with the vertical shear instability (VSI), one of the most promising hydrodynamic mechanisms, through 2D global non-ideal MHD simulations with ambipolar diffusion and Ohmic resistivity. 
For typical disk parameters, MHD winds can coexist with the VSI with accretion primarily wind-driven accompanied by vigorous VSI turbulence. The properties of the VSI remain similar to the unmagnetized case. The wind and overall field configuration are not strongly affected by the VSI turbulence showing modest level of variability and corrugation of midplane current sheet. Weak ambipolar diffusion strength or the enhanced coupling between gas and magnetic fields weakens the VSI. The VSI is also weakened with increasing magnetization, and characteristic VSI corrugation modes transition to low-amplitude breathing mode oscillations with strong magnetic fields. 

\end{abstract}

\section{Introduction}

Understanding the gas dynamics of protoplanetary disks (PPDs) is crucial for many aspects of planet formation. The key element to PPD gas dynamics lies in the mechanisms for disk angular momentum transport. This is firstly becuase it determines the overall disk structure and controls long-term disk evolution and secondly mechanisms responsible for angular momentum transport generally give rise to complex internal flow structures, particularly turbulence. Most stages of planet formation take place in PPDs, which sensitively depend on both of these factors (e.g., \citealt{armitage11}). While angular momentum transport is not yet directly observable, theoretical studies are mainly guided by the fact that PPDs are actively accreting onto protostars at a typical rate of $\sim10^{-8}M_\odot$ yr$^{-1}$ (e.g., \citealt{hartmann_etal98,hh08}), which places strong constraints on the possible mechanisms involved. 

Angular momentum transport in PPDs has been conventionally attributed to the turbulence driven by the magneto-rotational instability (MRI, \citealp{bh91}). However, this scenario is complicated by the fact that PPDs are extremely weakly ionized (e.g., \citealp{w07,bai11}), which substantially weakens the coupling between gas and magnetic fields. This coupling is described by three non-ideal magnetohydrodynamic (MHD) effects: Ohmic resistivity, Hall effect and ambipolar diffusion (AD). They affect the MRI in different ways (e.g., \citealp{w99,bt01,bb94,jin96,kb04,desch04}), and particularly dissipations by resistivity and AD lead to damping or even complete quenching of the MRI \citep{gammie96,bs11,simon_etal13a,simon_etal13b}. Via detailed simulations with increasing realistic disk microphysics, it has been shown that PPD accretion is likely primarily driven by magnetized disk winds \citep{bs13,bai13,gressel_etal15,bai17} in a largely laminar disk: a major shift of paradigm.

Parallel to the development of the MHD theory, pure hydrodynamic mechanisms for driving angular momentum transport have also been discussed extensively in recent years. These include the vertical sheear instability (VSI; \citealp{nelson_etal13}; hereafter N13), convective overstability in its linear \citep{kh14,lyra14,latter16} and non-linear state (subcritical baroclinic instability; \citealp{kb03,petersen_etal07a,petersen_etal07b,lp10}), and zombie vortex instability \citep{marcus_etal13,marcus_etal15,umurhan_etal16}. These instabilities generally operate under certain specific thermodynamic conditions, and can potentially be activated in different regions of PPDs \citep{fl17,malygin_etal17,pk19}. It is generally found in numerical simulations that the level of turbulence resulting from these hydrodynamic instabilities is not sufficiently strong to account for the typical disk accretion rates; they yield Shakura-Sunyaev \citep{ss73} $\alpha$ values up to $\sim 10^{-4}-10^{-3}$, while $\alpha\sim10^{-2}$ is likely required for the outer disk (e.g., \citealp{hartmann_etal98}). But as potential sources of disk turbulence, they have important implications to planet formation. They can efficiently stir up dust particles (e.g. \citealt{sk16,flock_etal17,lin19}), trap them in vortices and facilitate planetesimal formation (e.g. \citealt{raettig_etal15,mk18}). Such turbulence are also needed so that sub-micron dusts are suspended to account for the disk SEDs \citep{dalessio_etal01} and as seen in scattered light images. 

So far, studies of wind-driven accretion generally adopted simplistic treatment of thermodynamics, partly to avoid the development of these hydrodynamic instabilities. On the other hand, studies of these hydrodynamic instabilities mostly ignored magnetic fields. A natural question arises: are hydrodynamic and MHD mechanisms compatible with each other? In other words, can these hydrodynamic instabilities operate in disks that launch magnetized disk winds?

In this paper, we focus on one of the most promising hydrodynamic instabilities, the VSI, and its interplay with magnetic fields. The VSI is an application of the Goldreic--Schubert--Fricke instability \citep{gs67,fricke68}, originally derived in the context of differentially rotating stars, to accretion disks \citep{ub98,urpin03,au04}. Applications of the VSI to PPDs by N13 have drawn significant attention, leading to intense followup studies since then (e.g., \citealp{bl15,umurhan_etal16}). Essentially, a disk that is Rayleigh-stable from radial shear becomes unstable in the presence of vertical shear where the rotational velocity varies over height, provided that cooling timescale is much shorter than orbital timescale to overcome the stabilizing effect of vertical buoyancy. In the limit of instant cooling, the criterion becomes (N13)
\begin{equation}
\frac{\pa j^2}{\pa R}-\frac{k_R}{k_Z}\frac{\pa j^2}{\pa Z}<0,
\label{eq:criterion}
\end{equation}
where $j$ is the specific angular momentum and is a function of cylindrical coordinates $(R, Z)$, and $k_R, k_Z$ are the radial and vertical wave numbers. The VSI taps free energy from the vertical shear in disk rotational velocities, which destablizes inertial-gravity waves (e.g., \citealp{ly15}). Such vertical shear is present when the fluid is baroclinic where contours of constant density and pressure are not aligned, and this is generally unavoidable in passively heated disks. The requirement for rapid cooling or thermal relaxation is more demanding. For standard models of PPDs, calculations suggest that the VSI can be triggered in the outer regions beyond $\sim5$ AU as well as the very inner regions \citep{ly15,malygin_etal17,pk19}. 

Unstable VSI modes typically have large ratio of $k_R/k_Z$ to take the advantage of the vertical shear, and non-linear simulations by N13 in idealized setup assuming vertically isothermal disks have found that the VSI develops into vigorous turbulence with prominent vertical oscillations. Cooling time well within $\sim0.1$ orbital time is needed to trigger the VSI, giving $\alpha$ values up to $\sim10^{-3}$. More realistic simulations by \citet{sk14} incorporating radiative transfer show agreement with N13 results, though a reduced efficiency of angular momentum transport is reported. Recent 3D simulations further suggest the development of vortices (\citealp{richard_etal16,mk18}), which is likely due to secondary Kelvin-Helmholtz instabilities \citep{lp18}. 

In this paper, we conduct 2D global non-ideal MHD simulations of PPDs and study the possible development of the VSI, focusing on outer regions dominated by AD. While the VSI may operate in the innermost disk regions that are highly optically thick, the dynamics of these regions are much more complex and are likely to be dominated by the MRI turbulence due to the thermal ionization (e.g., \citealp{fromang_etal02,dt15,flock_etal17b}). We restrict ourselves to cases where the disk is otherwise laminar (without VSI) to avoid complications from external driving sources. Note that even the system is laminar, there is still finite coupling between gas and magnetic fields, and hence a pure hydrodynamic understanding is incomplete. Recently, \citet{lp18} analyzed the linear properties of the VSI in the presence of magnetic fields. They found that in ideal MHD limit, while the MRI and the VSI modes lie in the same branch of the dispersion relation, MRI modes always dominate given its much higher growth rate. It is speculated that the VSI modes might survive when the MRI is suppressed, though they only mentioned Ohmic resistivity. We will directly examine whether the VSI can survive in the AD-dominated regime more relevant to realistic PPDs. If so, we will further address how the VSI is affected by magnetic fields, and in turn, whether the development of the VSI affects wind properties. As the first numerical simulations on the interplay between the VSI and the non-ideal MHD, we aim to make the problem as clean as possible. Therefore, we do not employ sophisticated implementations of realistic disk physics as in \citet{bai17}, but rather conduct numerical experiments similar to those in \citet{bs17} with prescribed AD coefficients to mimic outer PPD conditions. We anticipate our results to serve as a benchmark for future studies that incorporate more realistic thermodynamics together with non-ideal MHD physics.

This paper is organized as follows. In \S\ref{sec:method}, we provide detailed descriptions of the numerical methods and the simulation setup. In \S\ref{sec:diag}, we list diagnostic quantities to furnish analyses of simulation results. Discussion on fiducial models of instability features are detailed in \S\ref{sec:result}. The flow structure, mass accretion, and disk winds are discussed in \S\ref{sec:wind}. We conduct parameter study on magnetic field strengths, cooling timescales, and ambipolar diffusion strengths in \S\ref{sec:parameter}. Finally, we summarize and discuss the main findings in \S\ref{sec:conclusion}.

\section{Methods and Simulation Setup}\label{sec:method}

In this section, we describe the method and setup of our simulations, and in the meantime demonstrate the basic considerations that enter these simulations.

\subsection{Dynamical equations}

We use the grid-based high-order Godunov MHD code Athena++ (Stone et al., submitted) to conduct global simulations of the VSI in the context of PPDs. Athena++ is the successor of the Athena MHD code \citep{gs05,gs08,stone_etal08}, but is rewritten in C++ which has much more flexible coordinate and grid options with significantly improved performance, scalability, and source code modularity. We solve standard MHD equations in spherical polar coordinates in conservative form, including non-ideal MHD effects as implemented in \citet{bs17}
\begin{gather}
\pdv{\rho}{t} + \div{(\rho \vb{v})} = 0, \\
\pdv{(\rho\vb{v})}{t}+\div{\left(\rho\vb{v}\vb{v}-\frac{\vb{B}\vb{B}}{4\pi}+\vb{P^*}}\right) = -\rho\grad{\Phi}, \\
\pdv{\vb{B}}{t}=\curl{(\vb{v}\times\vb{B}}-c\vb{E}^\prime), \label{eq:induc} \\
\pdv{E}{t}+\div{\left[(E+P^*)\vb{v}-\frac{\vb{B}(\vb{B}\cdot\vb{v})}{4\pi}+\vb{S}^\prime\right]} =-\rho(\vb{v}\cdot \grad \Phi) -\Lambda_\mathrm{c},  \label{eq:energy}
\end{gather}
where $\rho$, $\vb{v}$, $P$ are gas density, velocity, and pressure. The sum of the thermal and magnetic pressure is denoted by $P^*=P+B^2/8\pi$, where $\vb{P^*}=P^*\vb{I}$, with $\vb{I}$ being the identity tensor, and $B = |\vb{B}|$. The gravitational potential of the central star with mass $M_*$ has the form $\Phi=-GM_*/r$. The total energy density is given by $E = P/({\gamma-1})+\rho v^2/2+B^2/8\pi$, where $v = |\vb{v}|$. The ideal gas law is adopted with adiabatic index $\gamma=7/5$ for molecular gas in the bulk disk.\footnote{The VSI primarily occurs in the disk where gas is molecular with $\gamma=7/5$. By a transition above the disk surface, the gas is mainly atomic in the wind zone where $\gamma=5/3$ (e.g., \citealt{wang_etal19}). The exact value of $\gamma$ affects the critical thermal relaxation timescale to trigger the VSI \citep{ly15}.  Nevertheless, as we have experimented, we find effectively no difference in the outcome of the simulations between these two values.} The term $\Lambda_{\rm c}$ in the last equality represents the cooling rate to be detailed in \S\ref{sec:thermo}.

The electric field involves components from non-ideal MHD effects. In the local fluid rest frame, it reads
\begin{equation}
\vb{E}^\prime=\frac{4\pi}{c^2}{(\eta_O\vb{J}
+\eta_A\vb{J_\perp})},
\end{equation}  
where the Ohmic and ambipolar diffusivities are denoted by $\eta_O$ and $\eta_A$, and we have ignored Hall term (see Section \S\ref{sec:Elsasser}). The current density is $\vb{J}=c\curl\vb{B}/4\pi$, and we express $\vb{J_\perp} = - (\vb{J} \times \vu{b}) \times \vu{b}$ as the component of $\vb{J}$ that is perpendicular to the magnetic field. The magnetic field $\vb{B}$ has its unit vector denoted by $\vu{b}=\vb{B}/B$. Poynting flux associated with non-ideal MHD is given by $\vb{S}^\prime=c\vb{E}^\prime\times\vb{B}/4\pi$. We use Gaussian units in the above equations, whereas in code units a factor of $4\pi$ is absorbed so that magnetic permeability is $\mu=1$. 

We perform global 2D simulations in spherical polar coordinates ($r, \theta, \phi$). We also use $R=r\sin\theta$ and $z=r\cos\theta$ to denote radial and vertical components in cylindrical coordinates. The unit system in the code has $G=M=R_0=1$, where $R_0$ is the reference radius fixed at the location of the inner boundary. The simulation domain covers from $r=1$ to $100$, and from $\theta=0$ to $\pi$ (i.e., including the polar region) so that the domain contains sufficient dynamical range and is fully extended to accommodate wind launching. We use the van Leer time integrator, the HLLD Riemann solver, with piecewise linear reconstruction. Super time-stepping is used to accelerate the calculations of non-ideal MHD as in \citet{bs17}.

\subsection{Disk model and initial conditions}

As mentioned earlier, we aim to conduct relatively clean numerical experiments, hence our disk model is set to be scale-free. In particular, disk temperature is set by the aspect ratio $\epsilon= H/r$, where $H=c_s/\Omega$ is the disk scale height, and $c_s^2 = P/\rho$ is the isothermal sound speed. Being scale-free (self-similar) requires $\epsilon$ to be independent of $r$, so that disk temperature is given by
\begin{equation}
T=\frac{P}{\rho}=T_0\frac{R_0}{r}\epsilon^2(\theta)\ ,\label{eq:temp}
\end{equation}
where $T_0 = v_{\rm K0}^2=GM/R_0$. We consider disk temperature profile in $\theta$ to be
\begin{gather}
\epsilon(\theta) = \epsilon_{\mathrm{d}} + \frac{1}{2}(\epsilon_{\mathrm{w}} - \epsilon_{\mathrm{d}})\left[\tanh\left(\frac{\rm n\delta \theta}{\mathrm{\epsilon_{d}}}\right)+1\right].
\end{gather}
In the above, the aspect ratio morphs from a disk value $\epsilon_\mathrm{d}$ to a surface value $\epsilon_{\rm w}$. We employ a modestly thin disk with $\epsilon_\mathrm{d}=H_\mathrm{d}/r = 0.1$ and set $\epsilon_\mathrm{w} = 0.5$. The transition happens at $\theta_\mathrm{trans} = 3.5H_\mathrm{d}$, with $\delta \theta$ the angle from $\theta_\mathrm{trans}$. The transition width is controlled by $\mathrm{n}$, where in our prescription $\mathrm{n=2}$, corresponding to a transition occurs within $0.05H_\mathrm{d}$. Physically, the temperature transition is motivated by the external UV and X-rays heating from the protostar in the wind zone (e.g., \citealp{glassgold_etal04,walsh_etal10}).

We consider a density profile that is a power-law in radius 
\begin{gather} 
\rho = \rho_0\left(\frac{r}{R_0}\right)^{-q_D}f(\theta),  \label{eq:den}
\end{gather}
where we set $q_D=2$. The function $f(\theta)$ is obtained through hydrostatic equilibrium solution. Solving for the Euler's equation in spherical polar coordinates in $r$ and $\theta$ directions yields \citep{bs17}
\begin{gather}
\frac{d\ln F(\theta)}{d\ln \sin\theta} = \frac{GM}{T_0R_0}\frac{1}{\epsilon^2(\theta)}-(q_D+1), \label{eq:F} \\
v_\phi^2(r,\theta)=\frac{GM}{r}-(q_D+1)T_0R_0\frac{\epsilon^2(\theta)}{r}, \label{eq:vphi}
\end{gather}
where we define $F(\theta)\equiv f(\theta)\epsilon^2(\theta)$. Note that $f(\theta)$ is an implicit function of $\epsilon(\theta)$, hence numerical integration is applied to calculate $F(\theta)$ and then to compute $f(\theta)$. Radial and $\theta-$ velocities are set to zero for equilibrium, but we further add random noises to these velocity components with amplitude of $\pm5$ percent of the local sound speed.

The poloidal magnetic fields are initialized by specifying an azimuthal vector potential \citep{zanni_etal07,bs17},
\begin{equation}
A_\phi(r,\theta) = \frac{2B_{z0}R_0}{3-q_D}\left(\frac{R}{R_0}\right)^\frac{1-q_D}{2}[1+(m\tan \theta)^{-2}]^{-\frac{5}{8}}.
\end{equation}
Here, $B_{z0}$ is the midplane field strength at $R_0$. The parameter $m$ determines the initial magnetic field configuration, quantifying the degree for which the magnetic field lines bend. A value approaches infinity $m \rightarrow \infty$ gives a pure vertical field. In this paper, we choose $m=0.5$. The poloidal field is computed by $\vb{B}=\nabla \times \vb{A_\phi}$, so that one can arrive at $\vb{B}_\mathrm{mid}=B_{z0}(R/R_0)^{-(q_D+1)/2}\hat{z}$ for magnetic field at the midplane. Using vector potential guarantees that the resulting field is divergence free. Midplane poloidal field $B$ is set so that the plasma $\beta$, defined as the ratio of gas to magnetic pressure, is constant. With midplane $\beta_0 = 8\pi p_0/B^2_\mathrm{mid}$, $p_0$ being midplane pressure, we choose $\beta_0=10^4$ in the fiducial model and further explore stronger and weaker fields. 

\subsection{Els\"{a}sser numbers}\label{sec:Elsasser}

The strength of the non-ideal MHD effects are characterized by the dimensionless Els\"{a}sser numbers. For resistivity and ambipolar diffusion, they are given by
\begin{equation}
\Lambda=\frac{v_\textrm{A}^2}{\eta_O\Omega_\textrm{K}}, \quad Am=\frac{v_\textrm{A}^2}{\eta_\textrm{A}\Omega_\textrm{K}},
\end{equation}
where $v_\mathrm{A}=\sqrt{B^2/4\pi \rho}$ is the Alfv\'{e}n speed. At a given ionization fraction, $\eta_O$ is independent of field strength or density, whereas $\eta_A\propto B^2/\rho^2$, and the Hall diffusivity $\eta_H\propto B/\rho$. Note that by definition, $Am$ is generally independent of field strength. The Ohmic and Ambipolar Els\"{a}sser numbers are known to largely control operation of the MRI, with threshold about unity (e.g., \citealp{turner_etal07,bs11}). Here, we primarily consider the outer regions of PPDs where low densities make AD the dominant non-ideal MHD effect. While the Hall effect likely plays an important role at intermediate disk radii (e.g., \citealp{bai17}), we do not include it to avoid complications, and that it would break the scale-free nature of our numerical experiments. We do include resistivity for purely numerical reasons (see below).

In our fiducial simulations, we set $Am=0.3$ for the bulk disk. Note that the value of $Am$ is found to be on the order of unity towards PPD outer regions \citep{bai11,bai11b}, and we choose it to be on the lower end which helps suppress the MRI. The value of $Am$ then smoothly increases from the disk zone to the wind zone where $Am$ is set to $100$, attributed to the stellar irradiation of FUV and X-rays that substantially elevate the ionization level above the disk surface \citep{pbc11}, recovering the ideal MHD regime. The functional form of the transition is similar to that of temperature, since heating and ionization in the disk atmosphere/wind zone are both due to UV/X-rays.

For simulations with pure AD, the system unavoidably develops a current sheet in the midplane region where $B_\phi$ flips sign (e.g., \citealp{bs17,bai17,suriano_etal18}). This current sheet tends to be unstable, which then corrugates and develops into more complex structures (see Appendix \ref{app:current} for more discussion). While this is interesting on its own right and will be investigated in a future publication, it imposes difficulties to assess the development and saturation of the VSI and causes problems near the inner radial boundary. Given the experimental nature of our simulations, we thus apply resistivity near the inner boundary and a thin layer ($\sim\pm H_\mathrm{d}$) in the midplane region throughout to stabilize the system (see Figures \ref{fig:vert}). It allows us to investigate the development of the VSI on top of a laminar disk that launches MHD disk winds. The value of midplane resistivity is fixed to $\eta_\mathrm{O} = 0.05c_sH_\mathrm{d}$, and it rapidly declines towards the surface being negligible within $0.1H_\mathrm{d}$. The fiducial run with $\beta_0=10^4$ has an initial midplane Ohmic Els\"{a}sser number of $\Lambda=4\times10^{-3}$. 

\subsection{Thermal relaxation}\label{sec:thermo}

Cooling of the system is achieved through thermal relaxation, which is associated with the $\mathrm{\Lambda_c}$ term on the right hand side of Equation \eqref{eq:energy}. We adjust disk temperature as
\begin{equation}
\frac{dT}{dt} =-\frac{(T-T_\mathrm{eq})}{\tau},
\label{eq:coolfunc}
\end{equation}
where $T$ is the temperature at time $t$, and the target temperature $T_\mathrm{eq}$ is set to the equilibrium temperature by initial condition. A relaxation time $\tau$ is prescribed to be a fraction of the local Keplerian orbital period, i.e. $\tau(R)\propto \mathrm{P_{orb}}=2\pi/\Omega_\textrm{K}(R)$. More precisely, we adjust disk temperature at each timestep by
\begin{equation}
\Delta T =(T_\mathrm{eq}-T)[1-\exp(-\frac{\Delta t}{\tau})],
\end{equation}
where $\Delta T$ = $T(t+\Delta t)-T(t)$. The terminology utilized for thermodynamics throughout this paper is i) locally isothermal, which for $\Delta t \gg \tau$ the disk temperature is kept to its equilibrium value at each position; ii) thermal relaxation, for which disk temperature is recovered to its equilibrium value on some timescale determined by $\tau$.
Following \citet{bai17} and \citet{bs17}, the wind zone is set to be locally isothermal for all runs with a smooth transition from disk zone.\footnote{Global non-ideal MHD simualtions coupled with thermochemistry performed on disk winds indicate a $\tau \sim 5$ in the atmosphere \citep{wang_etal19}. More realistic thermodynamic prescription can be applied to the wind zone in the future work.}

\subsection{Boundary conditions}

The conditions of inner radial boundary deserve special attention. We fix the hydrodynamic variables to initial equilibrium state, with temperature and density computed from Equations \eqref{eq:temp} and \eqref{eq:den}. The angular velocity is set to the minimum between initial $v_\phi$ through Equation \eqref{eq:vphi} and $\Omega_\mathrm{K}(r_0)R$, along with the other two velocity components $v_r=v_\theta=0$. Such a fixed state boundary condition can provide more stable flow structure and minimize the influence from inner boundary to the main simulation domain. Furthermore, we add a buffer zone between $r=r_0$ and $r=1.5r_0$ by including a constant resistivity from the midplane all the way up to near the pole with values at each $\theta$ equal to midplane resistivity. It helps stabilize the magnetic flux originating from the inner boundary though not completely over long times.

The rest of the boundary conditions are standard. The hydrodynamic variables in the outer radial boundary are extrapolated from the last grid zone, assuming $\rho \propto r^{-q_D}$, $T \propto r^{-q_T}$, $v_\phi \propto r^{-1/2}$. Radial and meridional velocities $v_r$ and $v_\theta$ are copied directly from the last grid zone, except setting $v_r = 0$ when $v_r<0$. Magnetic field variables in the inner and outer ghost zones are determined via $B_r \propto r^{-2}$, $B_\theta \propto \mathrm{const.}$, $B_\phi \propto r^{-1}$. With $\theta$ domain reaching to the pole, we employ polar boundary conditions, where $\theta$ and $\phi$ components are reversed across the polar boundary. 

\subsection{Simulation runs}\label{sec:runs}

\begin{deluxetable}{l  c c c}
\tablewidth{0.45\textwidth}
\tablecaption{List of Simulation Models and Parameters} 
\tablehead{
\colhead{Run} &
\colhead{$\beta_0$} &
\colhead{$\mathrm{\tau/P_{orb}}$} &
\colhead{$Am$} 
}
\startdata
FidH & -- &  0 & --  \\
ResH & -- &  0 & --   \\ \hline
Fid & $10^4$ & 0 & 0.3  \\
B3 & $10^3 $ & 0 & 0.3 \\
B5 & $10^5$ & 0 & 0.3  \\ 
t-3 & $10^4$ & $10^{-3}$ & 0.3\\
t-2 & $10^4$ & $10^{-2}$ & 0.3  \\
t-1 & $10^4$ & $10^{-1}$ & 0.3 \\   
Am0.5 & $10^4$ & 0 & 0.5 \\
Am1 & $10^4$ & 0 & 1.0 
\enddata
\tablecomments{FidH and Fid are hydrodynamic and MHD fiducial simulations, respectively. $\rm P_{orb}$ and $\rm P_0$ denote local orbital period and innermost orbital period. Run FidH has runtime of $600 \rm P_0$ and B3 has runtime of $800\rm P_0$, and the rest models have runtime of $500\rm P_0$.}
\label{table:runs} 
\end{deluxetable}

Our simulation runs are listed in Table \ref{table:runs}, with three main physical parameters: the disk magnetization $\beta_0$, the thermal relaxation timescale $\tau$, and the disk ambipolar diffusion Els\"{a}sser number $Am$. For the fiducial run, we choose $\beta_0=10^4$, $\tau=0$ and ${\rm Am}=0.3$. Models by varying these parameters are considered with $\beta_0\in[10^3, 10^5]$, $\tau\in[0,0.1]$,
and ${\rm Am}\in[0.3, 1]$. To facilitate analyses, we also conduct simulations with $\tau=1$ that are otherwise identical to runs Fid, B3 and B5 but with $\tau=1$ where VSI is not expected to develop.

The simulation models listed in Table \ref{table:runs} are all carried out with the same resolution, 1536$\times$480 grid cells in $(r,\theta)$, except model ResH. The radial domain spans from $r=1$ to $100$ and has logarithmic spacing. Note that our radial domain is much wider than previous hydrodynamic simulations of the VSI. This is because to properly accommodate the MHD winds, domain size should generally be much larger than wind launching radius. Grid spacing in $\theta$ direction increases by a constant factor of 1.006 per grid cell from the midplane toward the two poles so that the resolution at the midplane is four times finer than that at the pole. This enables us to achieve a resolution of about 32 cells per $H_\mathrm{d}$ in $r$ and 32 cells per $H_\mathrm{d}$ in $\theta$ at the disk midplane. Run ResH is a hydrodynamic run with limited $\theta$ domain and no temperature transition for comparison with earlier works in literature. Its limited spatial domain spans $r\in[1,4]$ and $\theta\in[-5H_\mathrm{d}/r, 5H_\mathrm{d}/r]$. The resolution achieves 96 cells per $H_\mathrm{d}$ in $r$ and 108 cells per $H_\mathrm{d}$ in $\theta$ with uniform grid spacing. In Appendix \ref{app:res}, we demonstrate our adopted resolution is sufficient for numerical convergence.

For comparison, pure hydrodynamic simulations with the identical setup as in MHD runs are also conducted by the same set of thermal relaxation timescales. Note that these hydrodynamic simulations differ from existing simulations (e.g., those in N13) in a way that we cover much more extended $\theta-$domain with a temperature transition. We will see that most VSI activity is bounded by this temperature transition region even the disk atmosphere is set to have instantaneous cooling. 

\section{Diagnostics}\label{sec:diag}

We list relevant diagnostic quantities in this section to facilitate the analyses of simulation results.

\subsection{Kinetic energy}

One quantity of great interest is the perturbed energy, which is defined as the sum of volume-integrated meridional and radial kinetic energies nomarlized by the initial equilibrium angular velocity $\mathrm{v_{\phi0}}$ (N13):
\begin{equation}
\mathrm{KE} = \frac{\iiint \rho (v_r^2+v_\theta^2) r^2 \sin\theta  \; \dd r \dd \theta \dd \phi}{\iiint \rho v_{\phi 0}^2 r^2 \sin\theta  \; \dd r \dd \theta \dd \phi}. 
\label{eq:KE}
\end{equation}
We compute the kinetic energies in a box of $r\in[2,4]$ and $\theta\in[-3H_\mathrm{d}/r,3H_\mathrm{d}/r]$. The evolution of kinetic energies will be frequently invoked in result sections to facilitate comparisons of different simulation models.

\subsection{Stress tensors}

In steady state, the equation of angular momentum conservation in cylindrical coordinates reads 
\begin{equation}
\frac{\dot{M}_\mathrm{acc}v_\mathrm{K}}{4\pi} = \pdv{}{R} \left(R^2\int^{z_\mathrm{wb}}_{-z_\mathrm{wb}} dz\langle T_{R\phi} \rangle \right) + R^2 \langle T_{z\phi}\rangle|^{z_\mathrm{wb}}_{-z_\mathrm{wb}},
\label{eq:acc}
\end{equation}
where the angle brackets stand for temporal and spatial average. The net mass accretion rate on the left hand side of Equation \eqref{eq:acc} is given by
\begin{equation}
\dot{M}_\mathrm{acc} = -2\pi R\int^{z_\mathrm{wb}}_{-z_\mathrm{wb}} \rho v_R dz\ ,
\label{eq:macc}
\end{equation}
and two terms on the right associate to transport due to the radial and vertical components of the angular
momentum flux, corresponding to the $R\phi$ and $z\phi$ components of the stress tensor, are defined as  \citep{bh98}
\begin{equation}
T_{R\phi} =  \rho\delta v_R\delta v_\phi - \frac{B_RB_\phi}{4\pi}\ ,
\label{eq:stress}
\end{equation}
\begin{equation}
T_{z\phi}= \frac{-B_zB_\phi}{4\pi}\ . 
\end{equation}
The hydrodynamic and magnetic parts of the stress tensors are known as Reynolds ($T_\mathrm{Rey}$) and Maxwell ($T_\mathrm{Max}$) stress, respectively. For vertical transport, we retain only the Maxwell stress, because the Reynolds component, $\rho v_zv_\phi$ does not carry excess angular momentum from the disk.
The vertical height of the wind base is marked by $|z_\mathrm{wb}|=3.5H_\mathrm{d}$ which separates the disk and atmosphere. Empirically, it is the location where non-ideal MHD dominated disk zone transitions to ideal MHD dominated wind zone. For better statistics, we do it over a time interval from 200 to $300 \rm P_0$  as well as over a cylindrical radial extent $R\pm0.1$ around the radius of interest. 

Radial transport is generally mediated by turbulence, and is characterized by the the classic dimensionless $\alpha$ parameter \citep{ss73}, and at each location it is defined as  
\begin{equation}
\alpha_{R\phi} = \frac{\langle T_{R\phi} \rangle}{P}.
\label{eq:alpha}
\end{equation}
The role of  radial (viscous) transport on disk accretion is reflected in the vertically integrated $\alpha$, defined as 
\begin{equation}
\alpha=\frac{\int^{z_\textrm{wb}}_{-z_\textrm{wb}} T_{R\phi} \:dz}{\int^{z_\textrm{wb}}_{-z_\textrm{wb}} P \:dz},
\label{eq:alphaprime}
\end{equation}
and by assuming steady-state viscously driven accretion, it yields an accretion rate given by
\begin{equation}
\dot{M}_{\rm vis} \approx \frac{4\pi}{v_K}\frac{\pa (\alpha c_s^2R^2\Sigma)}{\pa R}\sim
(4\pi R)\alpha\epsilon_d^2\Sigma v_K\ ,
\label{eq:visacc}
\end{equation}
where $\Sigma = \int^{z_\textrm{wb}}_{-z_\textrm{wb}} \rho \:dz$ is the surface density. In the second relation, we replace $\pa/\pa R$ by $R^{-1}$ for an order of magnitude estimate.

Vertical transport of angular momentum is mediated by magnetized disk winds. Because of long lever arm, it is more efficient than radial transport by a factor $\sim R/H$, given $B_R$ and $B_z$ are on the same order of magnitude. Inside the disk, wind-driven accretion is achieved by a torque exerted by the Lorentz force, and in the thin disk limit, $B_z$ is approximately constant, the local accretion velocity given by \citep{w07,bs13}
\begin{equation}
- \frac{1}{2}\rho\Omega_\mathrm{k}v_R \approx - \frac{B_z}{4\pi} \pdv{B_\phi}{z}\ ,
\label{eq:massflux}
\end{equation}
which states that accreting velocity is proportional to the vertical gradient of $B_\phi$. 

\subsection{Wind kinematics}

The local mass loss rate per logarithmic radius of wind is formulated by  
\begin{equation}
\dv{\dot{M}_\mathrm{wind}}{ \ln R} = 2\pi R^2(\langle \rho v_z \rangle |_{z_\mathrm{wb}}+\langle -\rho v_z \rangle |_{-z_\mathrm{wb}}),
\end{equation}
where $\dot{M}_\mathrm{wind}(R)$ is the cumulative wind mass loss rate within radius $R$. An important diagnostic of the magnetized winds is the Alfv\'{e}n radius, which is the point along a filed line where the poloidal gas velocity $v_\mathrm{p}$ equals the local Alfv\'{e}n velocity $v_\mathrm{Ap} = B_\mathrm{p}/\sqrt{4\pi\rho}$. The wind mass loss rate is related to the mass accretion rate in the bulk disk by \citep{fp95},
 \begin{equation}
\xi = \frac{1}{\dot{M}_\mathrm{acc}} \dv{\dot{M}_\mathrm{wind}}{ \ln R} = \frac{1}{2} \frac{1}{(R_\mathrm{A}/R_\mathrm{wb})^2-1},
\label{eq:leverarm}
\end{equation}
where $\xi$ is called the ejection index, and the ratio $R_\mathrm{A}/R_\mathrm{wb}$ is often referred to the magnetic lever arm. Thus, the location of the Alfv\'{e}n point can provide a convenient measure of the wind mass loss rate. 

\subsection{VSI linear modes}\label{sec:theory}

General linear analysis on pure hydrodynamic VSI can be found in N13. 
With locally isothermal equation of state, and in the short wavelength limit, the VSI growth rate $\sigma$ is given by
\begin{equation}
\sigma^2 \sim 2\Omega_0 \frac{k_Z}{k_R}\frac{\partial v_\phi}{\partial Z} - \kappa^2_0 \frac{k^2_Z}{k^2_R},
\label{eq:shear}
\end{equation}
where $\Omega_{0}$ is the angular velocity of disk at center of the shearing box, and $\kappa_0$ denotes epicyclic frequency. This expression directly relates the vertical shear $\partial v_\phi/ \partial Z$ to the growth rate of the wave modes.
The fastest growing mode then satisfies 
\begin{equation}
\frac{k_Z}{k_R}=\frac{\Omega_0}{\kappa^2} \frac{\partial v_\phi}{\partial Z}\ ,
\end{equation}
with growth rate 
\begin{equation}
\sigma^2_\mathrm{max}=\frac{\Omega^2_0}{\kappa^2} \left(\frac{\partial v_\phi}{\partial Z}\right)^2\ .\label{eq:sigmax}
\end{equation}

Local linear analysis incorporating magnetism has been conducted by \citet{lp18} under the assumption of ideal MHD. For the VSI modes to fit into the disk, its wavelength should not exceed disk scale height, and this gives an upper limit to field strength
\begin{equation}
\beta \gtrsim q^{-2},
\label{eq:hp18}
\end{equation}  
where $\beta$ represents the ratio of thermal pressure to magnetic pressure, and $q$ is a measure of the vertical shear strength. Taking $q \sim H/R=0.1$, we have $\beta \gtrsim 100$ \citep{bl15}. Equations \eqref{eq:shear} and \eqref{eq:hp18} will later be involved to discuss why VSI weakens with increasing disk magnetization (\S\ref{sec:fs}).

The above mentioned linear analyses are performed under the local shearing sheet assumption in short wavelength limit. Vertically global analyses demonstrate the emergence of two types of modes, namely the body modes near the midplane which can be further categorized into breathing and corrugation modes, and the high latitude surface modes when applying no-flow boundary conditions (N13; \citealp{bl15,mp15,umurhan_etal16b}). The non-linear evolution of the VSI is eventually taken over by large-amplitude corrugation modes.

\section{Fiducial Models}\label{sec:result}

Starting with an analysis on fiducial hydrodynamic model FidH, we demonstrate its consistency to the limited $\theta$-domain simulation ResH (\S\ref{sec:FidH}). Then we present results from MHD simulations by discussing fiducial model Fid in detail to illustrate the main features of the VSI when incorporating magnetic fields (\S\ref{sec:feature}, \S\ref{sec:vertpro}, and \S\ref{sec:turbulence}). Analyses on model B3 and B5 are detailed in \S\ref{sec:fs}.

In presenting the results, we use time in units of the innermost orbital time $\rm P_0= {P_{orb}}(R_0)=2\pi$. If not otherwise noted, we analyze the data at radii between $R=2$ and $R=4$, and average the results from time 200 to 300$\rm P_0$ for model Fid and B5, and 600 to 700$\rm P_0$  for model B3. These time intervals are chosen after the VSI has fully saturated (see Figures \ref{fig:KE} and \ref{fig:tspace}).

\subsection{The hydrodynamic runs} \label{sec:FidH}

\begin{figure*}[ht]
\epsscale{1.2}  
\plotone{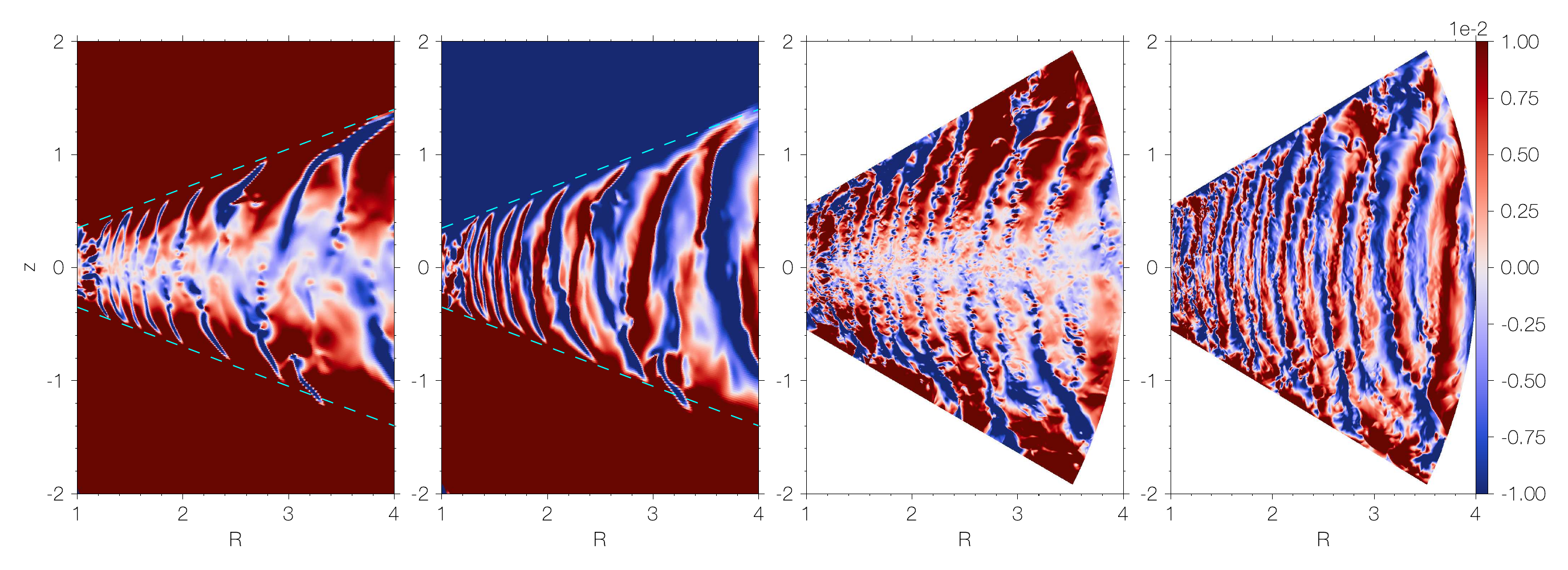}
\caption{Radial velocities $v_r$ for model FidH (first column) and ResH (third column) and meridional velocities $v_\theta$ for model FidH (second column) and model ResH (forth column) at $t= 150\rm P_0$. Dashed lines mark the angles $\pi/2-\theta=\pm0.35$ radian corresponding to the transition from disk zone to wind zone.}
\label{fig:FidH}
\end{figure*}

\begin{figure}[ht]
\epsscale{1.2}  
\plotone{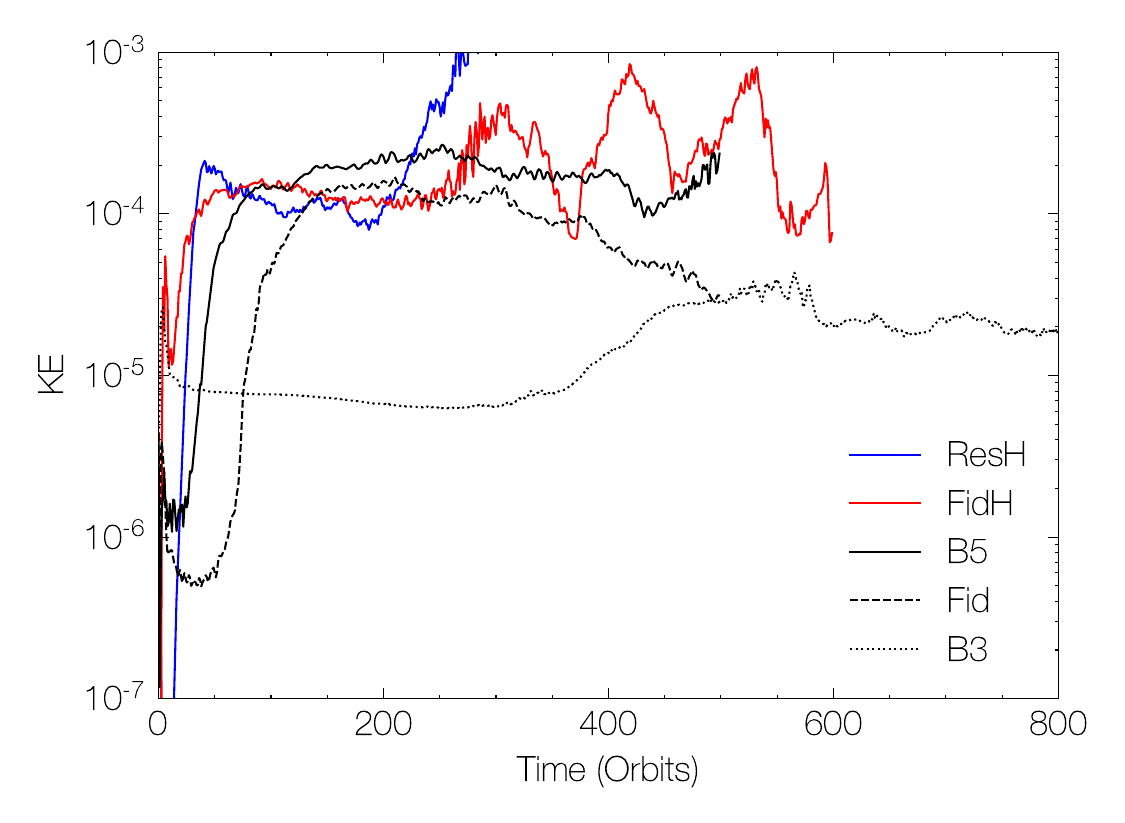}
\caption{Time evolution of normalized perturbed kinetic energies of model ResH, FidH, B3, Fid, and B5, all with locally isothermal prescription and plotted in log-linear scale.}
\label{fig:KE}
\end{figure}

In Figure \ref{fig:FidH}, we show the meridional velocities at time $t=150\rm P_0$ for run FidH and ResH. Note the difference between the two runs is that run FidH has an extended domain with a temperature transition at $z=\pm3.5H_\mathrm{d}$. As discussed in N13, there are higher latitude surface modes and lower latitude body modes in the early stage of their simulations, and the body (corrugation) modes dominate the non-linear state of evolution.
Surface modes generally require high resolution to be resolved due to their short radial wavelengths. Our run ResH achieves comparable resolution as in N13, and all these modes are observed where breathing modes show up when initial perturbations are small ($ \sim 10^{-6} c_s$). Surface and breathing modes are ultimately taken over by corrugation modes in the nonlinear stage, showing prominent vertical oscillations. While run FidH has lower resolution and does not show surface modes, we see that the corrugation modes at the non-linear state are well captured.

As seen in Figure \ref{fig:FidH}, the two runs share very similar turbulence patterns, except that in run FidH, the VSI is well confined in the bulk disk within the transition region. Turbulence properties in between the two runs also converge, where we find turbulent kinetic energy as defined in Equation \eqref{eq:KE} asymptotes to $\mathrm{KE} \sim 2\times10^{-4}$ after $\sim 40\rm P_0$ for both models (see Figure \ref{fig:KE}). The Reynolds stresses between the two models are also consistent with each other, on the order of $\alpha_{R\phi} \sim 10^{-3}$.

N13 found turbulent kinetic energy in their simulations rises unphysically after a few hundred orbits, which was attributed to their restricted computational domain and $\theta$ boundary conditions. We have also observed this phenomenon in run ResH. While our run FidH with much more extended domain shows fluctuations in kinetic energy after about $300\rm P_0$, the mean value remains similar, and the turbulence structure sustains in the long run.

\begin{figure*}[ht]
\epsscale{1.2}  
\plotone{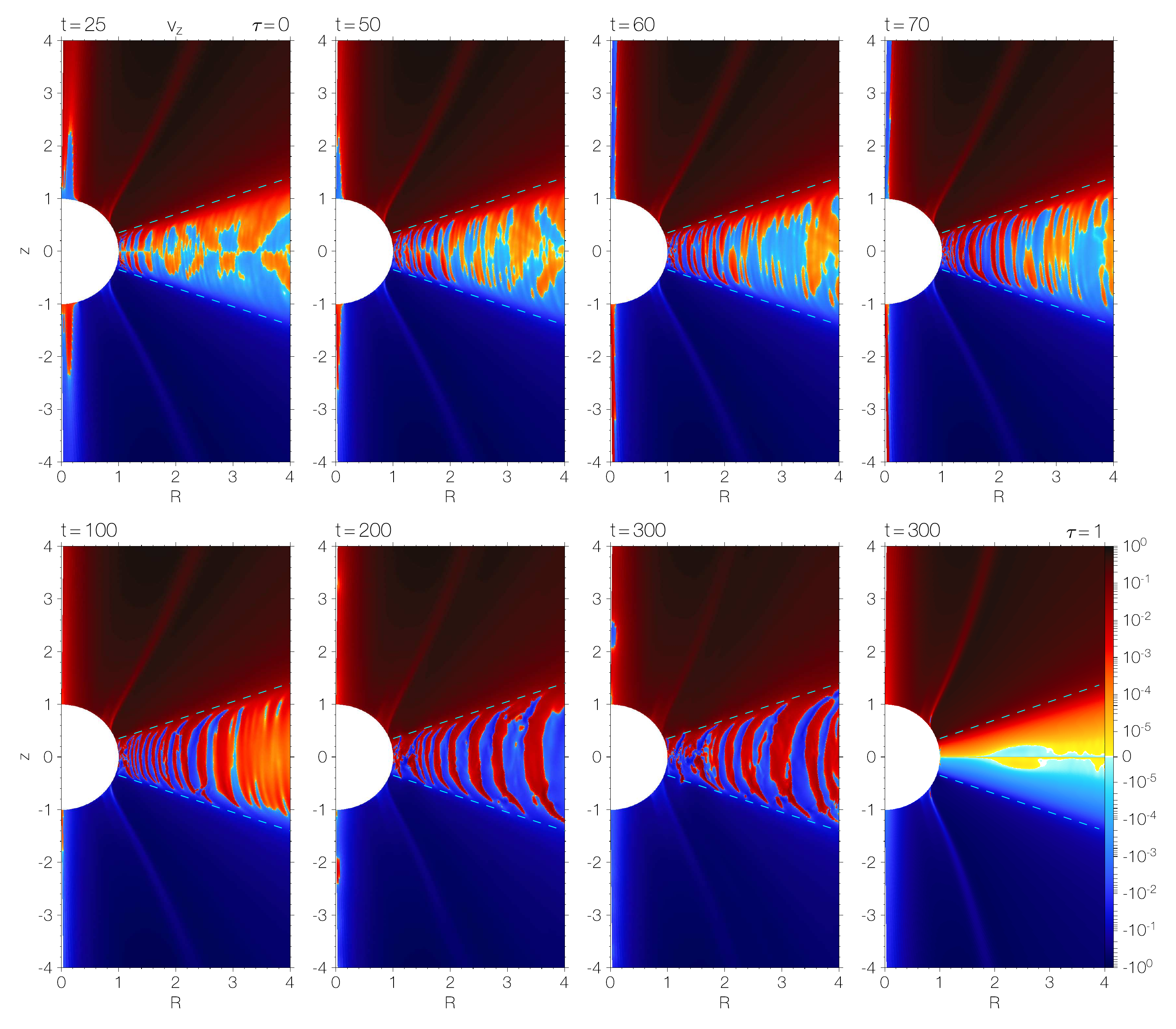}
\caption{Vertical velocities $\mathrm{v_z}$ displayed in logarithmic scale of the fiducial MHD simulation (model Fid). First seven panels: snapshots of vertical velocity at $ t=25,50,60,70,100,200$ and $300\rm P_0$ with locally isothermal prescription ($\mathrm{\tau=0}$). Lower right panel: snapshot of vertical velocity perturbations at $t=300\rm P_0$ with thermal relaxation prescription ($\mathrm{\tau=1}$). The dashed lines mark the opening angles $\theta=\frac{\pi}{2}\pm0.35$.}
\label{fig:fidvz}
\end{figure*}

\begin{figure*}[ht]
\epsscale{1.2}  
\plotone{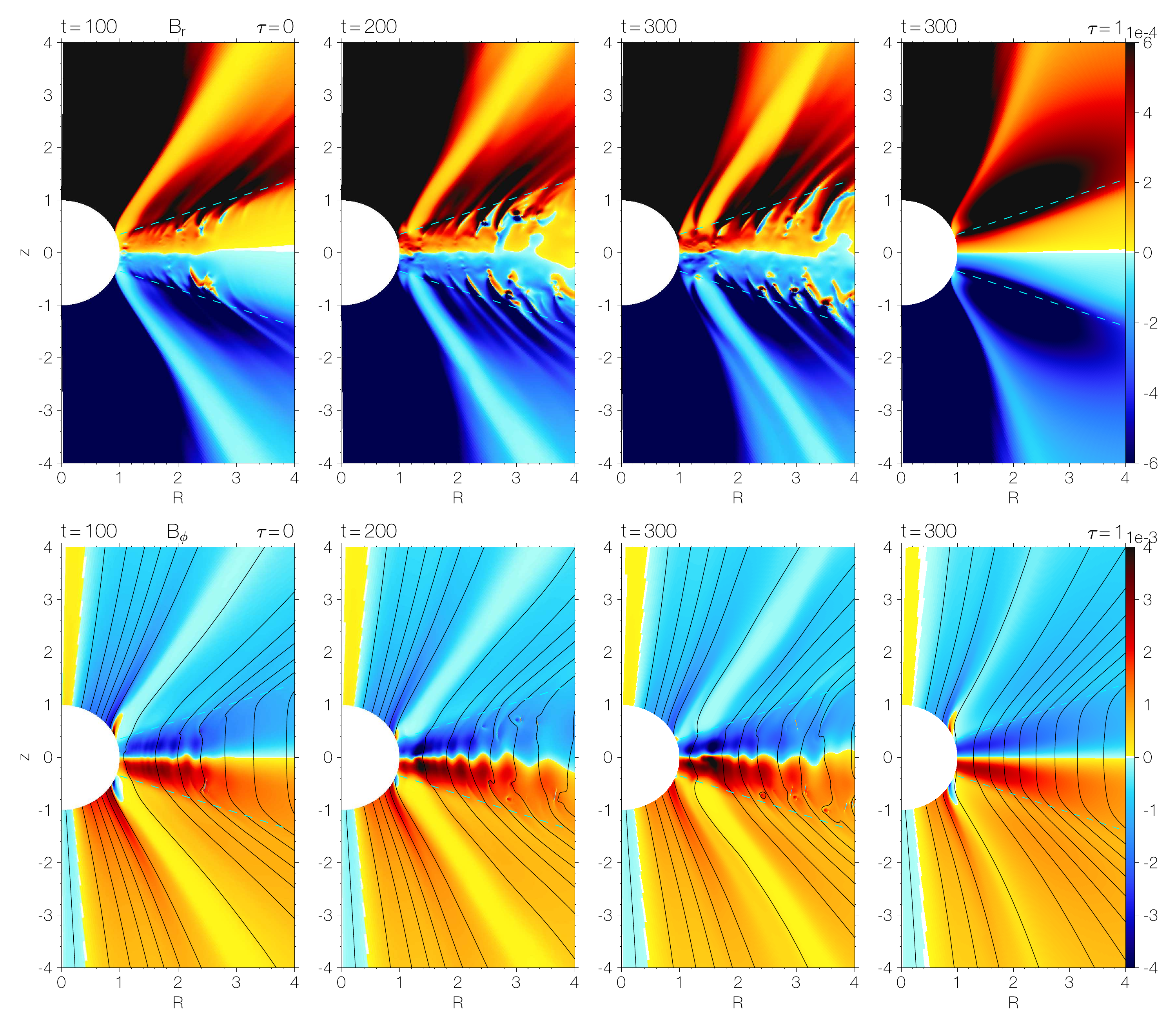}
\caption{Snapshots of radial (top panels) and toroidal (bottom panels) magnetic fields taken at $t=100,200$, and $300\rm P_0$  of fiducial model Fid. Solid black curves delineate the evenly spaced contour lines of poloidal magnetic flux. For comparison, the last columns of both rows show the corresponding $\tau=1$ contours at $t=300\rm P_0$. The dashed lines mark the opening angles $\theta=\frac{\pi}{2}\pm0.35$.}
\label{fig:fidB}
\end{figure*}

\begin{figure*}[ht]
\epsscale{1.2}  
\plotone{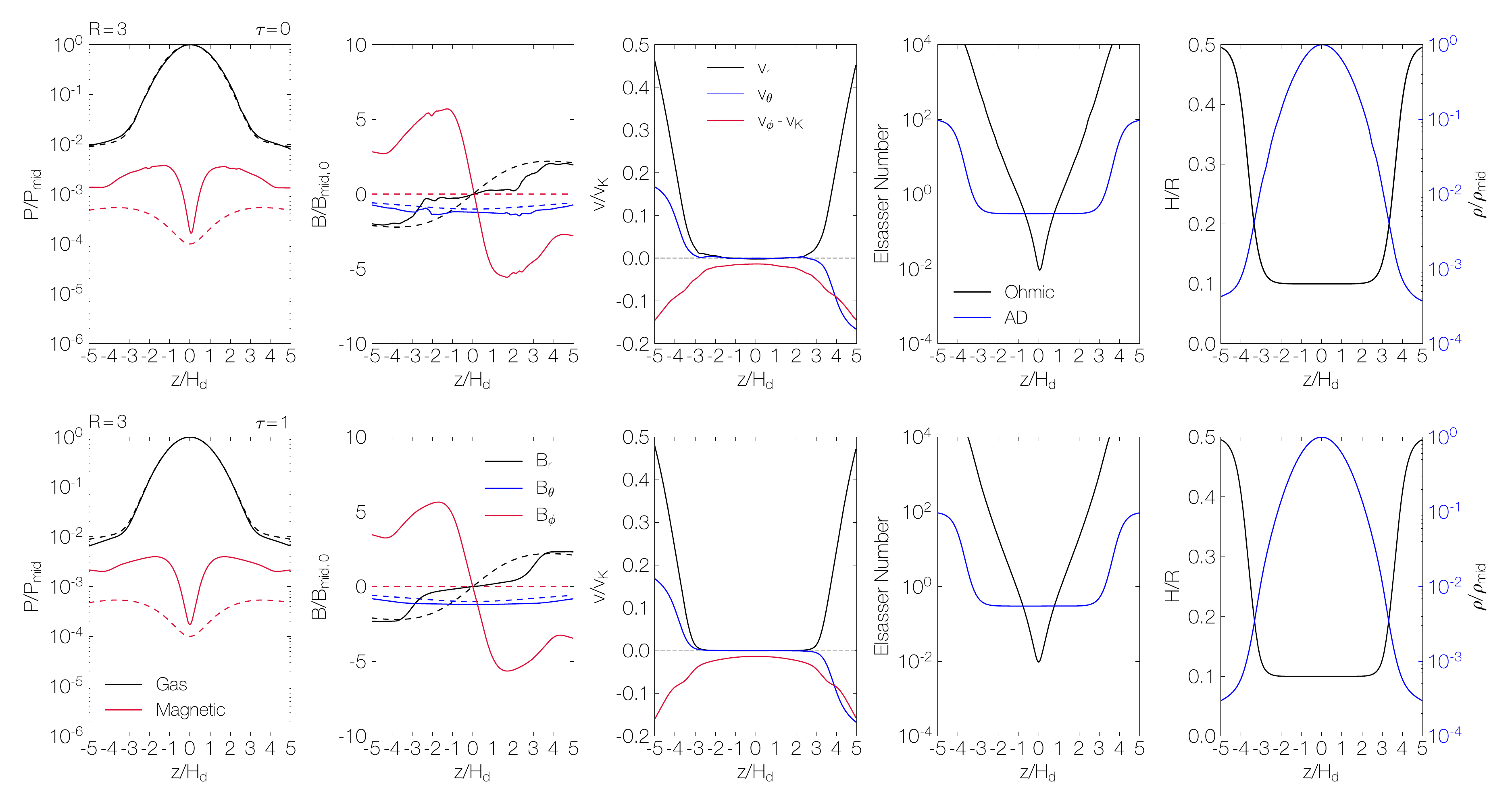}
\caption{Vertical profiles of hydrodynamic and magnetic variables at fixed cylindrical radius $R=3$ of MHD fiducial model Fid with initial magnetic field strength $\beta_0=10^4$. First row: profiles averaged over time interval from 200 to $300\rm P_0$ with thermodynamic prescription $\tau=0$. Second row: profiles measured at $t=300\rm P_0$ of corresponding $\tau=1$ model. Dashed curves correspond to profiles measured in initial hydrostatic equilibrium ($t=0$). Temperature profiles are shown in $H/R$. The gas density and pressure are normalized to midplane values $\rho_\mathrm{mid}$ and $P_\mathrm{mid}$. Magnetic field strength is normalized to initial midplane field strength $B_\mathrm{mid,0}$. Three velocity components are normalized to Keplerian velocity.}
\label{fig:vert}
\end{figure*}

\subsection{Time evolution in MHD simulations}\label{sec:feature}

The time evolution of our fiducial MHD run Fid is illustrated in four figures. Figure \ref{fig:KE} shows the time evolution of kinetic energy fluctuations defined in Equation \eqref{eq:KE}. In Figures \ref{fig:fidvz} and \ref{fig:fidB}, snapshots of vertical velocity $v_z$ and magnetic fields are displayed along the evolutionary sequence. In these two Figures, we also show results of the simulation with $\tau=1$ where the VSI does not develop for comparison. In Figure \ref{fig:tspace}, we provide a space-time plot of vertical velocity at $R=3$ to show the development of VSI consecutively.

With poloidal fields, initial stage of the evolution is characterized by wind launching from the surface layer, which occurs around the transition region where gas becomes better coupled to the magnetic field. This process generates some disturbances that traverse the simulation domain in a few tens of orbits that is visible in Figure \ref{fig:KE}. Once the wind is established in a few tens of $\rm P_0$, the overall field configuration remains quasi-steady state as seen in Figure \ref{fig:fidB}. Note that when setting the cooling time to be $\tau=1$ (last panels of Figure \ref{fig:fidvz} and \ref{fig:fidB}), the wind configurations remain stable and symmetric throughout the simulation time. Therefore, turbulent fluctuations we observe at later times in run Fid should arise from the VSI.

We observe the VSI being developed in the bulk disk body progressively from small to large radii, on top of the magnetized wind. In Figure \ref{fig:fidvz}, we see that at $ t=25\rm P_0$, wave patterns in a region $R<1.5$ are associated with VSI wave modes. Features at $R>1.5$ correspond to disturbances from initial relaxation. We mainly focus on regions between $R=2$ to $4$, for which it takes $\sim100 \rm P_0$ for the VSI turbulence to cover this entire region. 

The overall development of the VSI in run Fid is similar to that in the hydrodynamic counterpart run FidH. In Figure \ref{fig:KE}, we see the VSI of run Fid takes longer to develop, which is a general trend in magnetized runs (see \S\ref{sec:fs}). In Figure \ref{fig:fidvz}, it can be seen that regions around $R=2$ show  breathing modes (body modes with odd-symmetry) at $t=50\rm P_0$, and transitions at $t=60-70\rm P_0$ to become corrugation modes with even-symmetry. These features are all very similar to the pure hydrodynamic case, though the transition is slightly delayed when comparing the results at the same radial range. Due to the lower resolution, surface modes found in the initial development of the VSI in hydrodynamic simulations of N13 are not observed, though in N13 it was noted that these modes are later overtaken by corrugation modes. As can be seen in Figure 1, finer turbulent structures are present in higher resolution runs. It is unclear whether such finer turbulent structures affect the diffusion magnetic fields, though we expect them to be negligible compared to the much stronger non-ideal MHD effects. By $t=200\rm P_0$ and within the entire region of $R=4$, the instability is saturated with vigorous VSI turbulence, characterized by strong vertical oscillations. The radial spacing of these oscillation columns is about $\sim2-3H_{\rm d}$, again similar to the pure hydrodynamic case. Oscillation period can be also obtained in Figure \ref{fig:tspace}, which is about $20\rm P_0$ at $R=3$ corresponding to $3.8$ local orbits.

The development of the VSI modifies the magnetic field configuration. Around the transition zone, radial magnetic field appears to show some bunching behavior as seen in upper panels of Figure \ref{fig:fidB}. This results from radial shear in vertical oscillations of the VSI. In the disk zone, the $\tau=1$ simulation shows the standard pattern with smooth toroidal field that flips at the midplane, leading to a current sheet (e.g., \citealp{bs17}). In the bottom panels of Figure \ref{fig:fidB}, the current sheet corrugates and oscillates around the midplane in accordance with the VSI oscillations. As a consequence, the toroidal field component becomes less smooth showing some banded patterns.

One caveat in our simulations is that the magnetized wind configuration is only quasi-steady. As found in \citet{bs17}, disks tend to gradually lose magnetic flux over time. At numerical level, the inner boundary is gradually depleted of magnetic flux and becomes less magnetized, which causes the region to be less stable.
This can be seen in the sixth and seventh panels in Figures \ref{fig:fidB} and \ref{fig:fidvz}, and the effect starts to affect the region of interest after about 300$\rm P_0$, and is related to the successive decline of the kinetic energy beyond $\sim300\rm P_0$ seen in Figure \ref{fig:KE} and the corrugation oscillation period also becomes more irregular as seen in Figure \ref{fig:tspace}. Therefore, in the rest of the paper we restrict our analyses for model Fid in a time interval from 200 to $300\rm P_0$ and mainly focus at radius $R=3$.

\begin{figure*}[ht]
\epsscale{1.2}  
\plotone{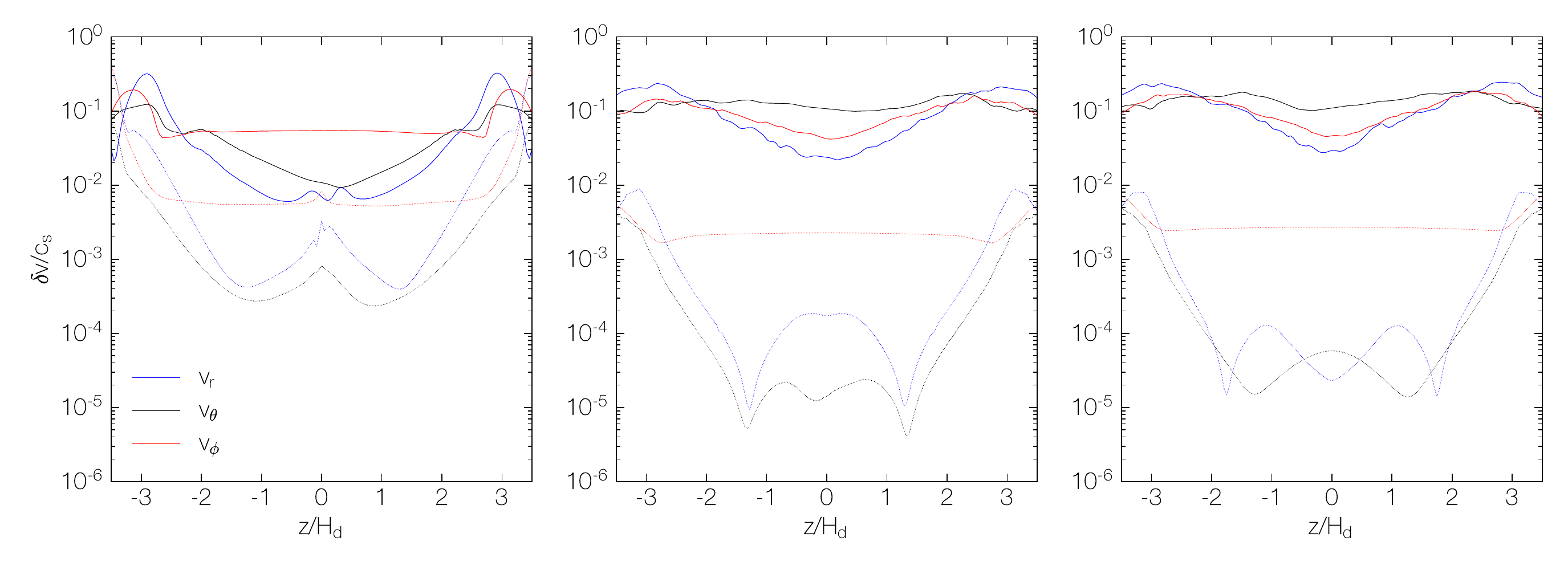}
\caption{Three components of velocity fluctuations normalized by local sound speed $c_s$ of model B3 (left column) averaged over $600-700\rm P_0$, Fid (middle column) and B5 (right column) averaged over $200-300\rm P_0$ at $R=3$. Solid and dotted curves correspond to $\tau=0$ and $\tau=1$ models, respectively.}
\label{fig:delta}
\end{figure*}

\subsection{Diagnostic vertical profiles}\label{sec:vertpro}

\begin{figure*}[ht]
\epsscale{1.2} 
\plotone{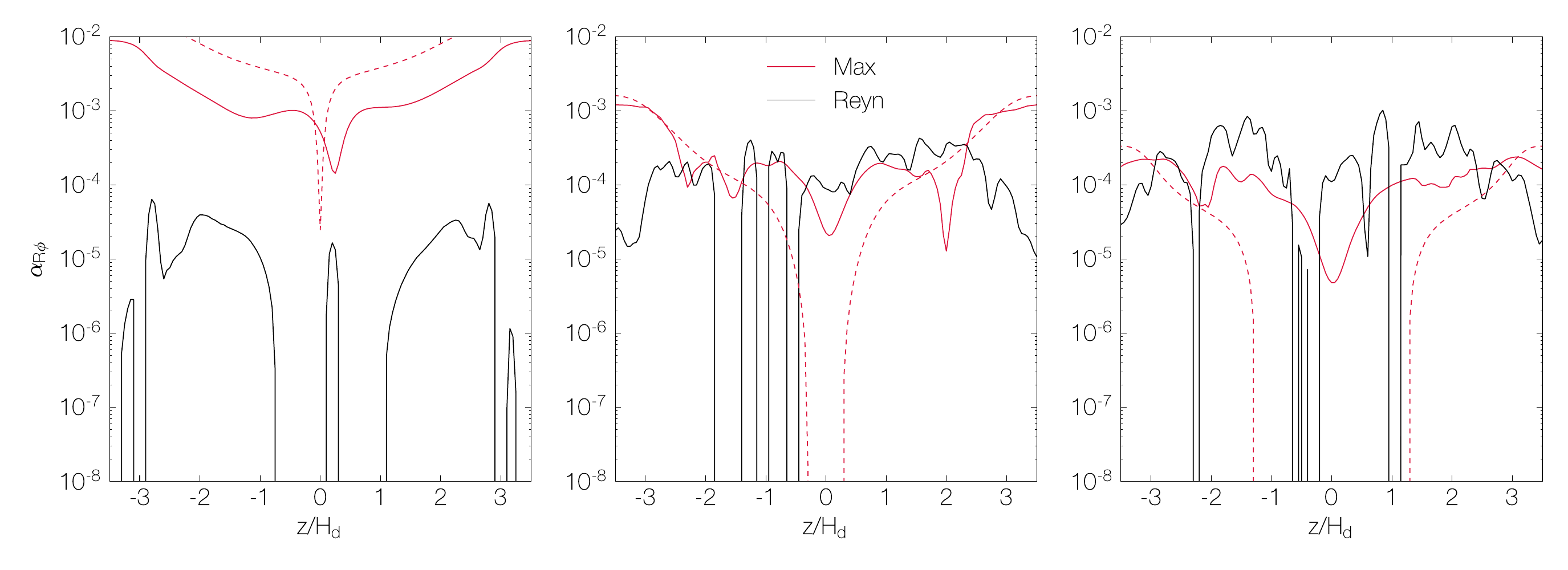}
\caption{Components of dimensionless $\alpha_{R\phi}$ parameter defined in Equations \eqref{eq:stress} and \eqref{eq:alpha}: averaged Reynolds (black) or Maxwell (red) stress normalized by the gas pressure at $R=3$ for model B3 (left panel), Fid (middle panel), and B5 (right panel). Solid curves delineate locally isothermal models $\tau=0$. Dashed curves delineate thermal relaxation models $\tau=1$.}
\label{fig:stress}
\end{figure*}

\begin{figure}[t!]
\epsscale{1.2} 
\plotone{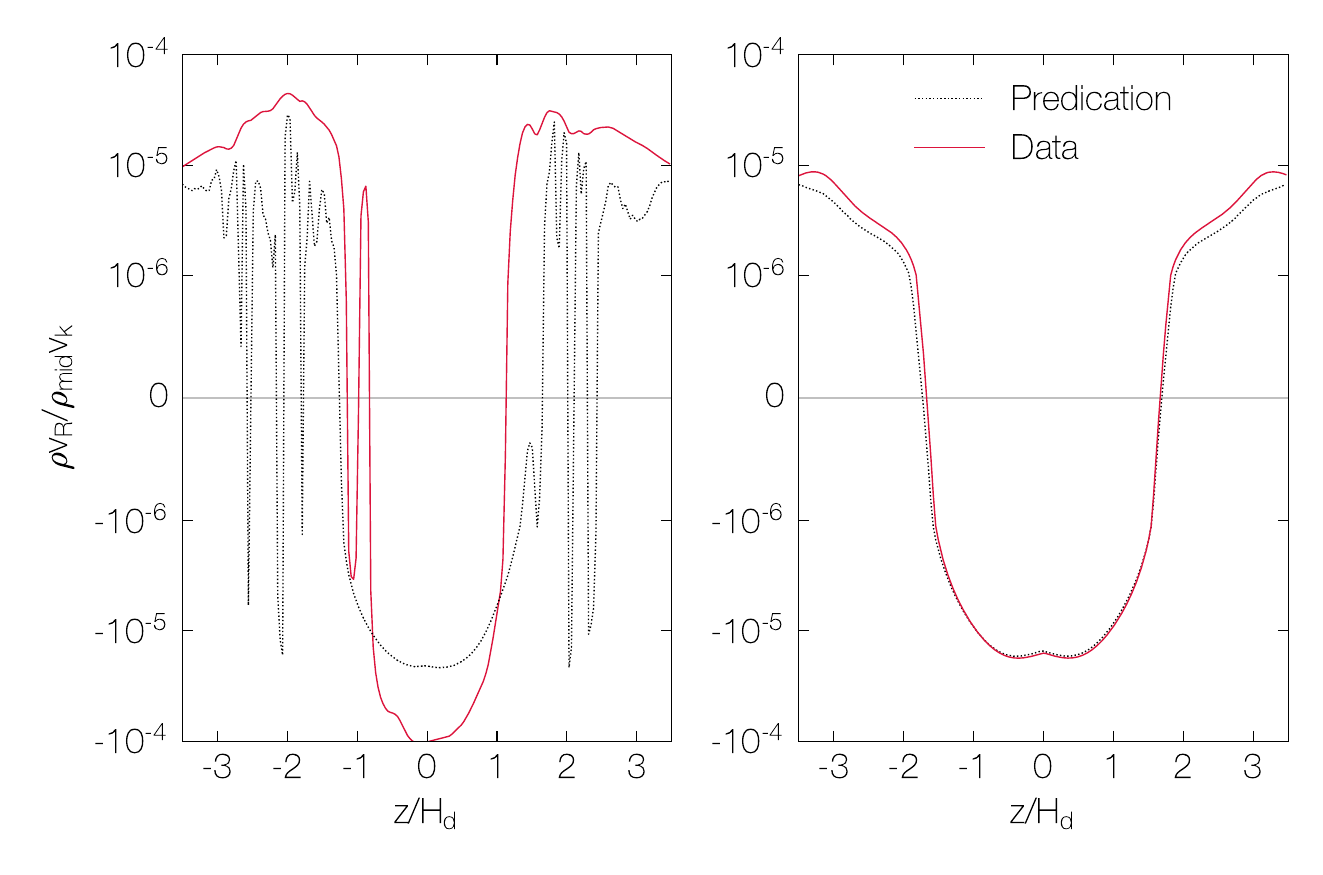}
\caption{Vertical profiles of mass flux measured at $R=3$ and averaged over $200-300\rm P_0$ of model Fid with $\tau=0$ (left) and $\tau=1$ (right). Mass fluxes computed directly from simulation data (solid red) and predicted from wind driven accretion (dotted black; Equation \ref{eq:massflux}) are shown. Grey curves delineate the transition from accretion to decretion region.}
\label{fig:rhovR}
\end{figure}

To examine the gas dynamics with the presence of the VSI, we show in Figure \ref{fig:vert} the vertical profiles of major diagnostic physical quantities in a vertical extent $z\in[-5H_\mathrm{d}/r,5H_\mathrm{d}/r],$ which covers the entire disk zone and part of the wind zone, at a cylindrical radius $R=3$. For comparison, the first and second rows show time averaged results from run Fid, and the corresponding profiles of the model with thermal relaxation time $\tau=1$. Dashed curves correspond to profiles measured in initial hydrostatic equilibrium ($t = 0$).

All profiles are either symmetric or anti-symmetric about the midpalne. In the second column, we see that toroidal fields are the dominant field component which reaches a factor $\sim5$ of midplane field and undergoes a flip of sign in the midplane. As seen in the fourth column,  the flip is achieved over a relatively thick layer of $z=\pm H$ due to the midplane resistivity added to stabilize the current sheet. Since the transition is at $z\sim\pm3.5H_{\rm d}$, the bulk disk is still AD dominated that is sufficient for us to study its interplay with the VSI for the outer disk. The relatively low transition height which may be expected for outer disk (e.g., \citealp{pbc11}) leads to the fact that the gas is still relatively dense there, and the entire disk is gas pressure dominated. Nevertheless, overall wind profiles remain similar to those obtained in previous wind simulations (e.g., \citealp{bs17}). The poloidal velocity is accelerated along field lines reaching a substantial fraction of midplane Keplerian speed, whereas rotation remains sub-Keplerian at all heights as seen in the third column due to the pressure gradient. A vertical shear in the rotational velocity is clearly present which is the driving source of the VSI. The wind profiles between run FidH which is turbulent but time averaged and its $\tau=1$ counterpart which is laminar without VSI closely resemble each other. This is a clear indication that the VSI does not affect the general properties of the disk wind.

\subsection{Turbulence strengths}\label{sec:turbulence}

To assess the level of VSI turbulence, we calculate the root mean square of velocity fluctuations as $\delta v\equiv (v-\langle v \rangle)_\mathrm{rms}$, where angle brackets denote the temporal and spatial averaging. We average the quantities over a time interval from 200 to $300\rm P_0$ and over a spatial extent $R=3\pm0.1$. The results are shown in Figure \ref{fig:delta} for runs B3, Fid, and B5.

We focus on run Fid here in the middle panel. The perturbed velocities in $\tau=0$ run is around $10\%$ local sound speed in the bulk disk, and is dominated by the vertical component as is expected in the VSI. The radial and azimuthal component of the fluctuation rises towards the wind zone. At the wind base around the transition height ($z\sim\pm3.5H_{\rm d}$), total velocity fluctuations is at about $20-30\%$ midplane sound speed, which leads to fluctuations in wind kinematics (see more discussions in \S\ref{sec:diskwind}). By contrast, velocity fluctuations are at much lower level by several orders of magnitude in the $\tau=1$ case.

\section{Angular Momentum Transport}\label{sec:wind}

In this section, we analyze angular momentum transport processes in presence of both the VSI and disk winds through the radial (\S\ref{sec:radialt}) and vertical (\S\ref{sec:vertt}) transport.

\subsection{Radial Transport}\label{sec:radialt}

\begin{figure*}[ht]
\epsscale{1.15} 
\plotone{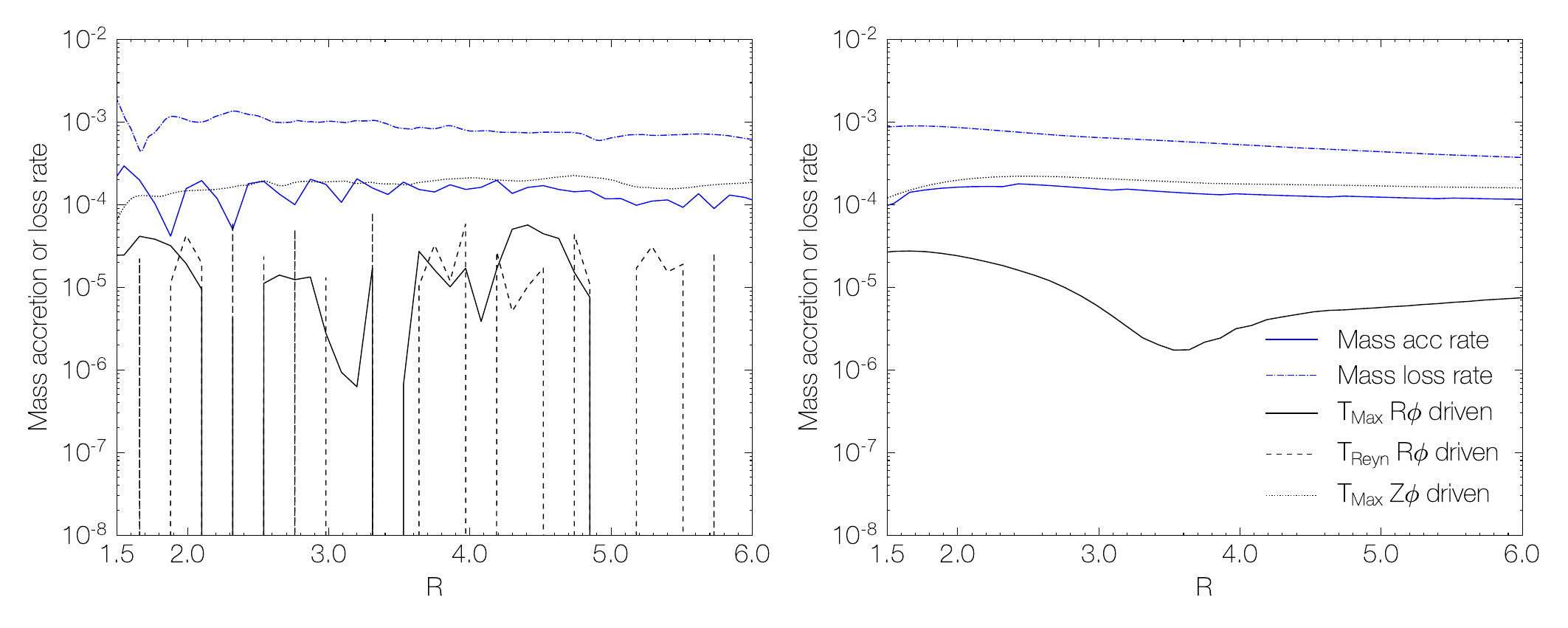}
\caption{Mass accretion and mass loss rate of model Fid (left panel) and corresponding $\tau=1$ model (right panel). The radial profiles of mass accretion rate (blue solid), mass loss rate per logarithmic radius (blue dash-dotted), $R\phi$ component of Maxwell stress driven accretion rate (black solid), $R\phi$ component of Reynolds stress driven accretion rate (black dashed), and $z\phi$ component of Maxwell stress driven accretion rate (black dotted) are shown.}
\label{fig:mdot}
\end{figure*}

\begin{figure}[ht]
\epsscale{1.2} 
\plotone{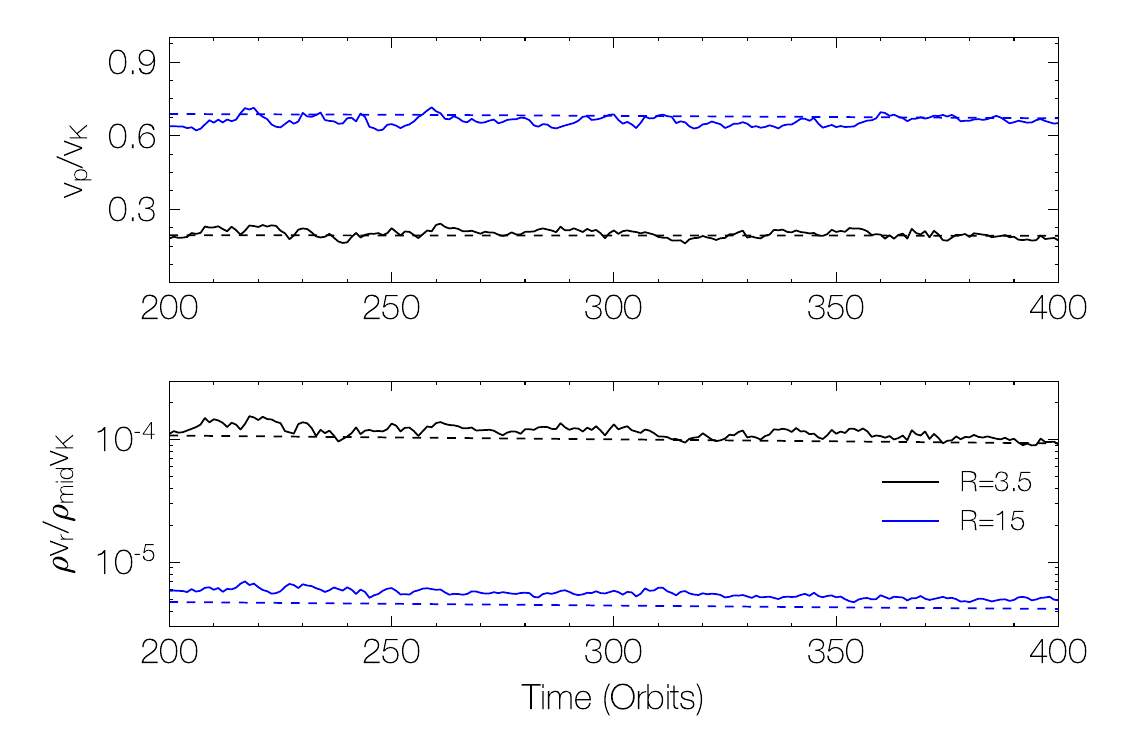}
\caption{Time variability of quantities measured at $R=3.5$ (black) and $R=15$ (red) along the traced poloidal magnetic field line from midplane with $R_0=3$. Top panel: time variability of wind poloidal velocity nomarlized by Keplerian velocity $\Omega_\mathrm{K}(R_0)$. Bottom panel: time variability of mass flux normalized by midplane density $\rho_\mathrm{mid}(R_0)$ and Keplerian velocity. Solid and dashed curves represent models of $\tau=0$ and $\tau=1$.}
\label{fig:timevar}
\end{figure} 

In Figure \ref{fig:stress}, we show components of $\alpha_{R\phi}$, the time and spatially averaged Reynolds and Maxwell stress, normalized by the midplane gas pressure at $R=3\pm0.1$. Focusing on the middle panel for our fiducial model Fid, the Reynolds stress due to the VSI is appreciable throughout the disk column achieving a maximum of $\sim 5\times 10^{-4}$ at $z\sim \pm 1.5H_\mathrm{d}$. The $\tau=1$ model shows a negligible Reynolds stress as expected.

The Maxwell stresses of both the locally isothermal model and thermal relaxation ($\tau=1$) model drop near the disk midpane ($z\sim \pm H_\mathrm{d}$,  Figure \ref{fig:vert}). While this is mainly due to (artificially) enhanced Ohmic resistivity applied to this region, we note that in reality the flip of $B_r$ and $B_\phi$ should give rise to a similar but much narrower feature. The Maxwell stress rises towards disk surface as accompanied by wind launching, and reaches  a local $\alpha_{R\phi} \sim 10^{-3}$ near the wind base. With the VSI, additional turbulent fluctuations increase the Maxwell stress in the midplane region so that the overall vertical profile is relatively flat compared to the $\tau=1$ case. 

When integrating the stress over height, one obtains the vertically integrated $\alpha$ of Maxwell and Reynolds stress. For model B3, Fid, and B5, the Reynolds stresses obtained are $\alpha \approx 6.9\times 10^{-6}, 1.0\times10^{-4}, 1.6\times 10^{-4}$, and the Maxwell stresses are $\alpha \approx 4.1\times 10^{-3}, 5.7\times 10^{-4}, 1.0\times10^{-4}$, respectively. With decreasing disk magnetization, the Reynolds stress becomes prominent, and it surpasses Maxwell stress as seen in model B5. 

We note that the Reynolds stresses measured in all three of our magnetized models are noisy and are smaller than those obtained in 3D simulations ($\alpha\sim10^{-3}$; e.g. N13; \citealp{mk18}) which allow for better averaging. We find the same holds for hydrodynamic simulations. The higher Reynolds stress in 3D is likely due to the formation of vortices \citep{richard_etal16,mk18}, though the kinetic energy fluctuations remain very similar between 2D and 3D. Further study with non-ideal MHD in 3D is needed to better constrain the range of Reynolds stress.

\subsection{Vertical Transport}\label{sec:vertt}

Without VSI turbulence, previous works (e.g., \citealp{bai17}) show that magnetized disk winds are predominant in driving disk accretion. In this section, we examine how the VSI affects the winds properties.

\subsubsection{Flow Structure}

As explained earlier, for purely wind-driven accretion, the vertical profile of accretion (radial) velocity should be given by Equation \eqref{eq:massflux}. This is clearly seen in the right panel of Figure \ref{fig:rhovR} for simulations with $\tau=1$, similar to the results in \cite{bs17}. The accretion layer is confined to within $z=\pm2H_\mathrm{d}$ where $B_\phi$ flips, and the mass flux agrees very well with expectations. Gas starts to flow radially outward as the gradient of $B_\phi$ reverses. Some deviation in the upper layer can be attributed to the fact that the $B_z$ can no longer be taken to be constant as assumed.

In the left panel of Figure \ref{fig:rhovR}, the onset of the VSI modifies flow structure in both accretion and decretion layers. The accreting/decreting layer becomes narrower/broader, and the mass fluxes of these two layers are both enhanced. The bulk disk is now not in steady state, hence Equation \eqref{eq:massflux} is inapplicable to estimate the flow structure. Sudden drops of the dotted black curve near $z=\pm 2.5 H_\mathrm{d}$ are associated with the bunching of poloidal fields, and VSI turbulence tends to restrict the accreting flow toward the midplane within $z=\pm H_\mathrm{d}$.

\begin{figure*}[ht]
\epsscale{1.2}  
\plotone{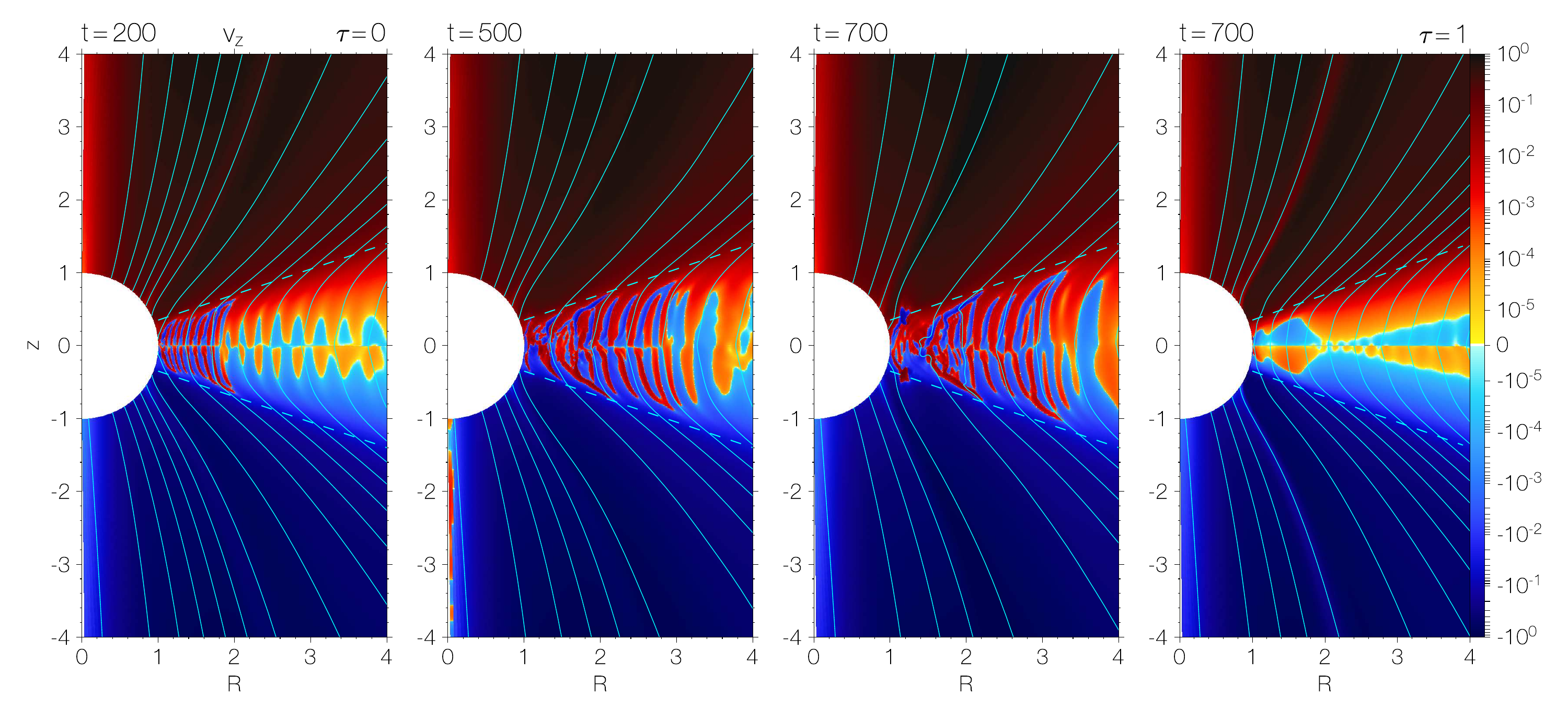}
\caption{Snapshots of vertical velocity $\rm v_z$ in logarithmic scale for model B3 at $t=200,500$ and $700\rm P_0$ for $\tau=0$ as well as at $t=700\rm P_0$ for $\tau=1$. Solid cyan curves delineate the evenly spaced contour lines of poloidal magnetic flux. The dashed lines mark the opening angles $\theta=\frac{\pi}{2}\pm0.35$.}
\label{fig:B3vz}
\end{figure*}

\begin{figure*}[ht]
\epsscale{1.2}  
\plotone{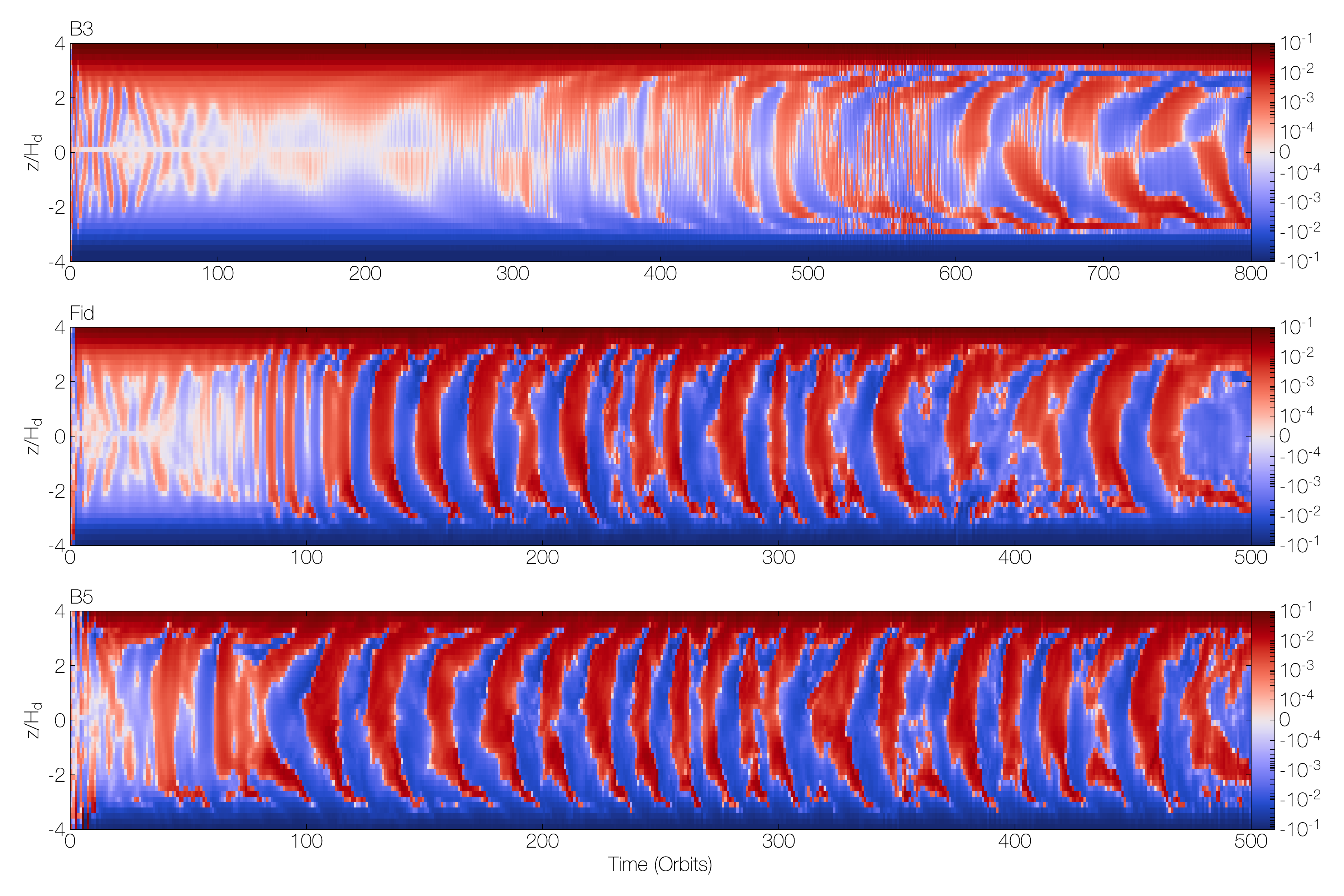}
\caption{Space-time diagram of vertical velocity $\rm v_z$ at $R=3$ for run B3 (top panel), Fid (middle panel), and B5 (bottom panel) in logarithmic scale. Note that the time is units of $\rm P_0$ (orbital period at the innermost boundary), which needs to be divided by $3^{3/2}\approx5.2$ to convert to local orbits at $R=3$.}
\label{fig:tspace}
\end{figure*}

\begin{figure}[ht]
\epsscale{1.25} 
\plotone{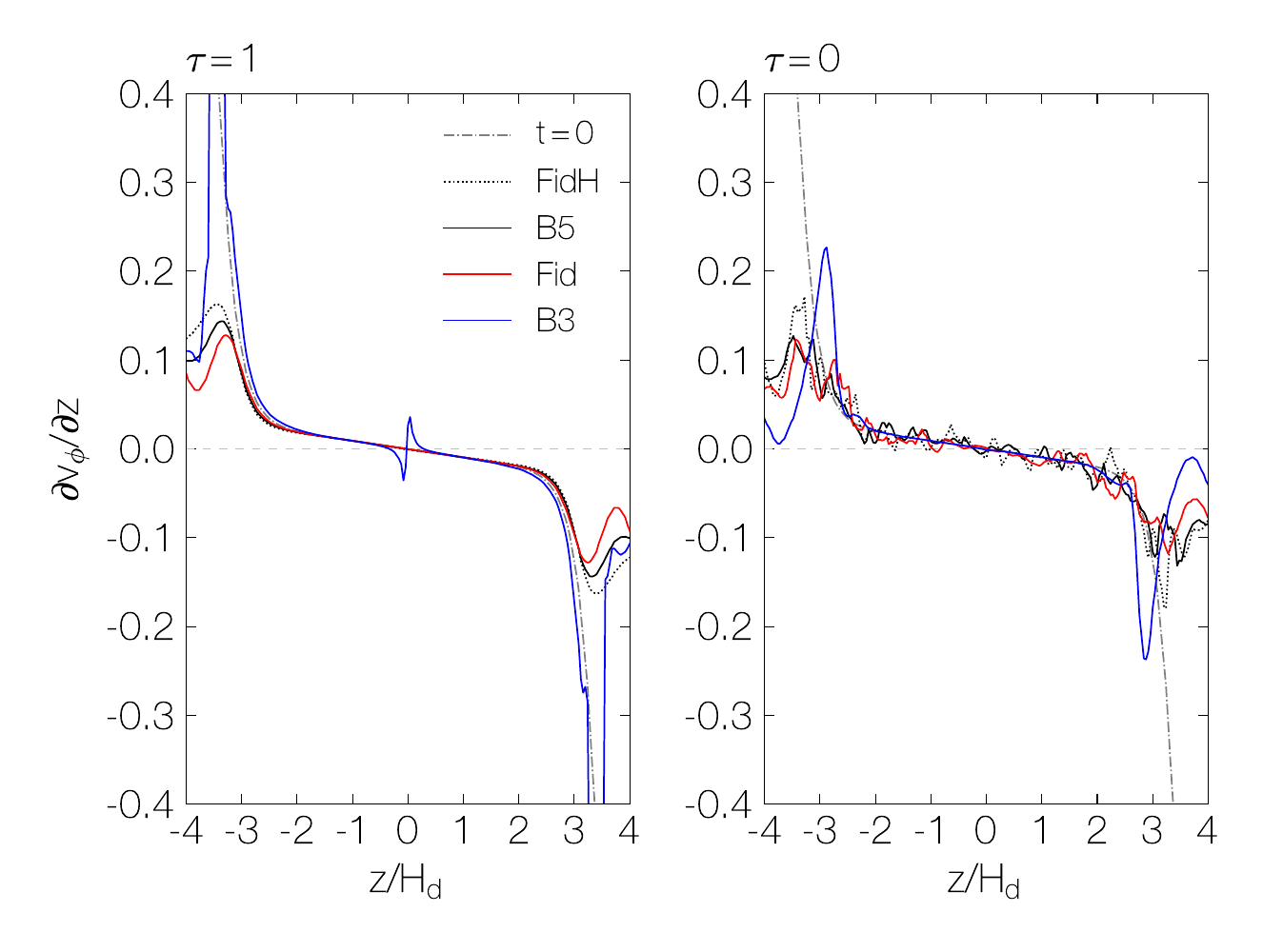}
\caption{Left: vertical shear profile at $R=3$ of model FidH (dotted), B5 (black), Fid (red) at $200\rm P_0$ and B3 (blue) at $600\rm P_0$ with $\tau=1$. Right: vertical shear profile at $R=3$ of model FidH, B5, Fid averaged over $200-300\rm P_0$ and B3 averaged over $600-700\rm P_0$ with $\tau=0$. In both panels, the dash-dotted curves display the initial vertical shear profile $(t=0)$, which are the same for all runs. The grey line marks zero vertical shear gradient.}
\label{fig:shear}
\end{figure}

\subsubsection{Mass accretion rate}

Figure \ref{fig:mdot} shows the radial profiles of accretion and mass loss rates, computed at $R=3$ and averaged over $t=200-300\rm P_0$. The right panel shows accretion rate without the VSI. Clearly, accretion is mainly driven by wind, and it dominates the contribution from $R\phi$ component of the Maxwell stress by a factor of $\sim R/H$, as expected. There is excessive mass loss, with mass loss rate a factor of several higher than accretion rate. Correspondingly, we find that the Alfv\'{e}n radius is very close to the wind base, as expected from Equation (\ref{eq:leverarm}). This is mainly controlled by the location of wind base, and higher wind base would lower the mass loss rate (e.g., \citealp{bai_etal16}). Detailed understanding of wind dynamics requires realistic calculations of heating/cooling process in the disk atmosphere (e.g., \citealp{wang_etal19}) and is beyond the scope of this work. Here, we are mainly concerned the gas dynamics in the bulk disk and the interplay between VSI and winds, and we have experimented to confirm that our overall conclusions are robust when setting the wind base at different heights.

The left panel of Figure \ref{fig:mdot} shows accretion rates in the presence of the VSI. The bulk accretion and mass loss rates are similar to the $\tau=1$ case, indicating that the VSI plays a minor role in the overall processes including angular momentum transport and mass loss. Still, magnetized wind remains the dominant mechanism in driving disk accretion. The main differences due to the VSI is that accretion driven by the $R\phi$ component of Reynolds stress is significantly boosted, reaching a level comparable to that from the Maxwell stress. We also note that with the VSI the profiles are noisier even after some time-averaging. To better quantify the role of radial transport, we use Equation (\ref{eq:visacc}) for an order-of-magnitude estimate, taking $\alpha\sim10^{-3}$, which is slightly larger than the values obtained from our fiducial run (sum of Maxwell and Reynolds stress), but comparable to those measured in 3D hydrodynamic simulations (\S\ref{sec:radialt}). Converting to code units, we find a value of $\sim1.8\times10^{-5}$ at $R=3$, which is again about a factor $H/R$ to wind-driven accretion rates.

\subsubsection{Wind variability}\label{sec:diskwind}

The turbulent nature of the bulk disk due to the VSI induces time variabilities in the wind. To show this, we trace a poloidal field line from midplane at $R_0=3$ all the way up to the boundary of the simulation domain. In Figure \ref{fig:timevar}, we present the time evolution of poloidal velocity and mass flux along this field line at cylindrical radii of $R=3.5$ and $R=15$ over a time interval $200-400\rm P_0$. The poloidal velocities measured at the two radii increases with distances, implying the wind is kept accelerating. The wind from the $\tau=1$ case (no VSI) is fairly steady, as the bulk disk is in quasi-steady-state. With the VSI ($\tau=1$), wind velocity and wind mass flux are on average about the same as the the $\tau=1$ case. In the meantime, wind velocity fluctuates at about $\pm 25\%$ at $R=3.5$ and $\pm 8.5\%$ at $R=15$, whereas the mass flux fluctuates around $\pm 34\%$ at $R=3.5$ and $\pm 26\%$ at $R=15$. These results suggest that the VSI affects wind kinematics only at a modest level.

\section{Parameter Study}\label{sec:parameter}

In this section, we investigate the effect of magnetic field strength (\S\ref{sec:fs}), the thermal relaxation timescale (\S\ref{sec:thermalrelax}), and the ambipolar diffusion strength (\S\ref{sec:Am}) on the onset of the VSI. 

\subsection{Magnetic field strength}\label{sec:fs}

\begin{figure*}[ht]
\epsscale{1.2} 
\plotone{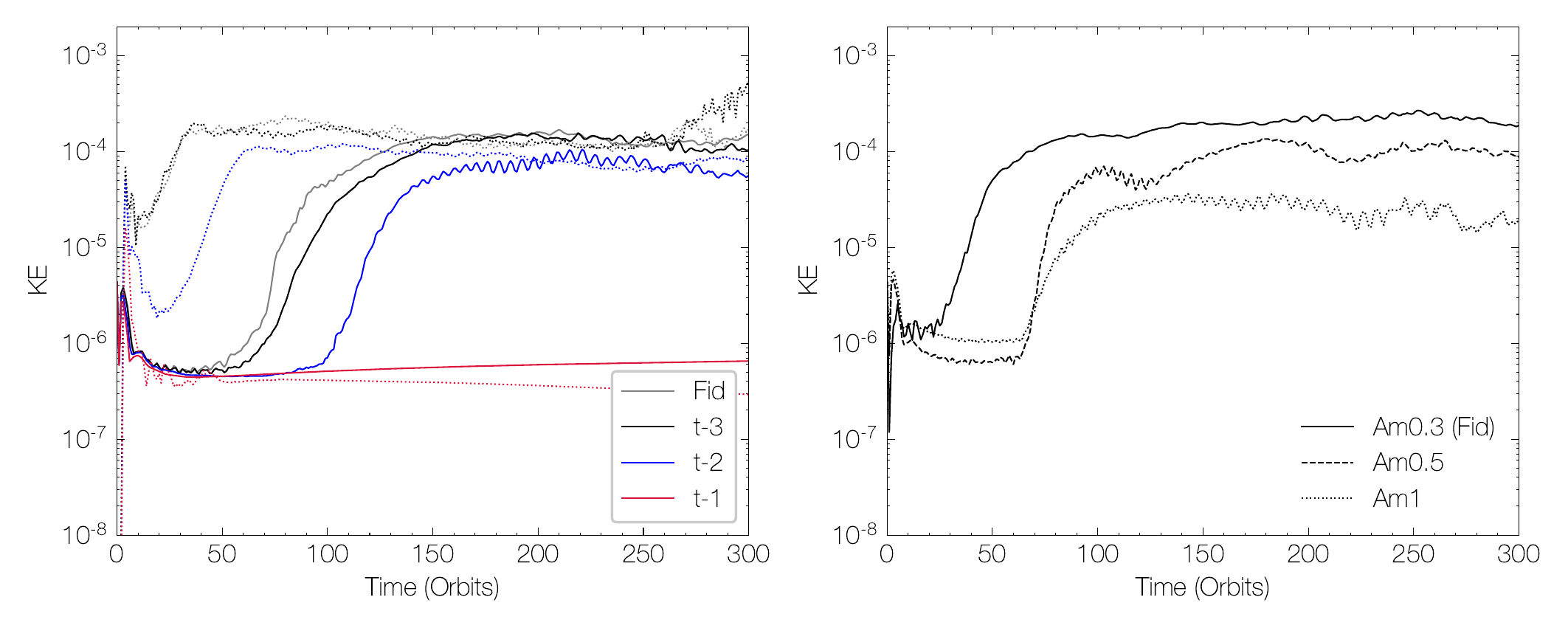}
\caption{Left panel: evolution of normalized kinetic energy with various thermal relaxation timescales. Solid curves show kinetic energies of model Fid ($\tau=0$), t-1, t-2, t-3. Dotted curves show kinetic energies of the corresponding hydrodynamic simulations to MHD models. Right panel: evolution of perturbed kinetic energy with varying ambipolar diffusion strengths of model Fid (Am=0.3), Am0.5, and Am1. Model FidH is also shown for comparison.}
\label{fig:kepara}
\end{figure*}

We first look at the evolution of kinetic energies for all runs in Figure \ref{fig:KE}. It is notable that the development of the instability is closely related to the magnetic field strength; stronger fields result in a time delay on the onset of the VSI and weaker kinetic energy fluctuations. Specifically, initial growth proceeds the fastest in the hydrodynamic run FidH. Growths in models B5 and Fid lag a few tens of orbital periods behind. These runs saturate at similar turbulence levels. The growth of model B3 is much more prolonged, and is only noticeable after $\sim 300\rm P_0$ and saturates after about $450\rm P_0$. Overall, run B5 resembles the behavior as run Fid except that the body (corrugation) modes show earlier growth and saturate to slightly higher amplitude (Figure \ref{fig:KE}). Run B3 is more special and we discuss in more detail below.

The vertical velocity fluctuations at different stages of the disk evolution for run B3 are shown in Figure \ref{fig:B3vz}. At $t=200\rm P_0$, we can identify the presence of breathing modes with odd symmetry. The amplitude of such fluctuations saturates after about $500\rm P_0$ at radii of $R=2-4$ and remains till the end of the simulation at $800\rm P_0$ (Figure \ref{fig:KE}). We find that the system is entirely dominated by such low-amplitude breathing mode (top panel of Figure \ref{fig:tspace}). The oscillation period at $R=3$ is around $\sim50P_0\sim10$ local orbits, and is not entirely steady as the radial spacing of such mode also varies with time. Because of the symmetry of the breathing mode, the current sheet remains largely unperturbed in the equatorial plane. We also show results from the simulation with $\tau=1$ in the last panel of Figure \ref{fig:B3vz}. By contrast, there is very little fluctuation in vertical velocity and no sign of such breathing mode, which further supports that what we observe in run B3 is a consequence of the VSI.

Figure \ref{fig:delta} further shows the vertical profiles of velocity fluctuations of run B5 and B3. We see that the velocity fluctuations of run B5 generally resemble the behavior of run Fid, which are dominated by vertical motions similar to the pure hydrodynamic case at quantitative level. Velocity fluctuations in run B3 is much weaker. Moreover, as the corrugation mode is suppressed, vertical velocity fluctuation appears to be the weakest, whereas there are stronger velocity fluctuations in $v_\phi$.

Figure \ref{fig:stress} is similar to Figure \ref{fig:delta}, but for vertical profiles of the $R\phi$ component of Reynolds and Maxwell stresses. For Maxwell stress, the trend is clear that stronger magnetization gives higher Maxwell stress which accompanies wind launching as usual.\footnote{In addition, midplane resistivity weakens the Maxwell stress more significantly for runs with weaker magnetization.} On the other hand, run B5 yields stronger Reynolds stress than the fiducial run, while run B3 gives much less. Since the Reynolds stress is mainly caused by the VSI, this anti-correlation with field strength again reflects the VSI weakens with stronger magnetization.

The fact that stronger magnetization leads to weaker VSI may have two possible causes. First, wind launching modifies the vertical shear profile and thus likely alters the free energy source that feeds the VSI. This can be seen in Equation (\ref{eq:sigmax}), where linear growth rate directly scales with this shear rate, though it does not apply to the vertically global body modes. In Figure \ref{fig:shear}, we show the vertical shear profiles of runs FidH, B5, Fid, B3 in $\tau=1$ (no VSI) and $\tau=0$ (with VSI) cases. Comparing the two panels, we see that the VSI modifies the shear profile at modest level. Interestingly, some narrow feature is observed at $\sim3H_\mathrm{d}$ in run B3 that is related to wind launching, in both $\tau=1$ and $\tau=0$ cases, whereas below this region, the shear profile remains similar to runs with weaker magnetization. Given the complexity of this feature, however, it is difficult to deduce whether it is responsible for the transition to breathing modes or not. 

The perspective above employs the results of linear analysis from pure hydrodynamics on top of a magnetically-modified vertical shear profile, thus ignores the coupling between gas and magnetic field. In the opposite limit, linear analysis by \citet{lp18} mainly assumed ideal MHD limit (i.e., perfect coupling between gas and magnetic field), and found that the MRI generally overwhelms the VSI, and that the VSI modes are stabilized by magnetic tension when $\beta\lesssim q^{-2}\sim100$ (Equation \ref{eq:hp18}). We find that magnetization in our run B3 has reached this limit for $z \geq 2H_\mathrm{d}$, which is consistent with the lower-than-expected growth rate discussed earlier. We will further discuss in \S\ref{sec:Am} on the role played by ambipolar diffusion. We also note that the linear analysis in \citet{lp18} was conducted in the short-wavelength limit, applicable mainly to the surface modes, whereas the more relevant is the body modes requiring vertically-global treatment. While this has been done in the hydrodynamic framework (N13; \citealp{bl15,mp15}), it can be highly challenging to further incorporate vertical magnetic fields because of the lack of a well-defined initial equilibrium.

\subsection{Thermal relaxation timescale}\label{sec:thermalrelax}

The thermal relaxation timescale is decisive to the onset of the VSI. Short thermal timescales can diminish the stabilizing effect from buoyancy, hence furnish the growth of the instability. Here, we consider the evolution of models by relaxing the locally isothermal assumption. The thermal timescales investigated are fractions of the local orbital periods. We chose $\tau=0.1,0.01,0.001$ corresponding to model t-1, t-2 and t-3, along with the fiducial model Fid where $\tau=0$. For comparison, each MHD model is accompanied by a hydrodynamic run with the same thermal relaxation prescription.

In Figure \ref{fig:kepara}, we show the evolution of the kinetic energy fluctuations for models with various thermal timescales. In pure hydrodynamic simulations, it is clear that in the $\tau=0.1$ model, the kinetic energy fluctuations quickly damps and then maintain some constant and very low level, indicating no VSI growth. Model $\tau=0.001$ shares great similarities with locally isothermal run FidH, whereas Model $\tau=0.01$ grows slower and saturates at a turbulence level with kinetic energy about twice lower. The critical cooling time of $\tau=0.1$ is consistent with previous analytical studies and hydrodynamic simulations with either simplified disk models or more realistic radiative transfer (e.g. N13; \citealp{sk14,ly15}).

The solid curves in Figure \ref{fig:kepara} show the corresponding MHD runs. Comparing the results with the pure hydrodynamic runs, we see that the threshold thermal relaxation timescale is similar between the two cases. For $\tau=10^{-3}-10^{-2}$, the level of saturation is also similar between hydrodynamic and MHD runs, where the system shows prominent vertical oscillations which are characteristic of the corrugation modes. 

\subsection{Ambipolar diffusion strength}\label{sec:Am}

The role played by non-ideal MHD effects (especially ambipolar diffusion) in our simulations can be understood from two aspects. First, they suppress the MRI, thus allowing the VSI to stand out. Second, they break the flux freezing condition to make the gas behave close to the unmagnetized case. \citet{lp18} qualitatively discussed the VSI under the Ohmic resistivity, where the VSI should be able to operate completely unimpeded when magnetic diffusion is able to weaken magnetic tension over the timescale of VSI growth, which translates to an Ohmic Els\"{a}sser number $\Lambda < q$, where $q$ is defined in Equation \eqref{eq:hp18}. Ambipolar diffusion shares some similarities with Ohmic resistivity in its dissipative nature, though a more detailed examination is needed. Here, we show the influence of $Am$ from our numerical results below.

We explore the evolution of perturbed kinetic energies as a function of ambipolar diffusion Els\"{a}sser number $Am$ which is prescribed as a constant within the bulk disk (Figure \ref{fig:vert}). Realistic values of $Am$ is found to be on the order of unity towards outer regions of PPDs (e.g., \citealp{bai11}).
Here, we further increase the value of $Am$ to 0.5 and 1, and the time evolution of kinetic energy fluctuation of these runs are shown in the right panel of Figure \ref{fig:kepara}. We see that the VSI becomes weaker when $Am$ is larger, which corresponds to smaller diffusivity. This result supports the expectation discussed earlier that strong magnetic dissipation can assist the development of the VSI.

From these results, we see that in the limit of strong ambipolar diffusion ($Am\ll1$), the disk can be considered as unmagnetized and we expect vigorous VSI turbulence if thermal relaxation in the disk is sufficiently rapid. In the opposite limit ($Am\gg1$), the MRI will take over the VSI that again leads to strong turbulence. Hence, weakest turbulence is likely to occur in between these two limits. 

\section{Conclusions and Discussion}\label{sec:conclusion}

In this paper, we study the onset and non-linear evolution of the vertical shear instability, which has so far been studied mostly in the hydrodynamic framework, in the presence of MHD disk winds. We perform 2D simulations in spherical polar coordinates in the $r-\theta$ plane using Athena++ code. We focus on outer regions of PPDs, where thermodynamic conditions favor the development of the VSI. Non-ideal MHD effects including ambipolar diffusion and Ohmic resistivity are incorporated, with prescribed diffusivities that suppress the MRI in the bulk disk and launch MHD winds from the surface. The main results are summarized as follows.

\begin{itemize}
\item The VSI generates vigorous turbulence in the presence of magnetized PPD disk winds. The turbulence properties are similar to pure hydrodynamic case, yielding velocity fluctuations on the order of $0.1c_s$ dominated by vertical motions. The Shakura-Sunyaev $\alpha$ parameter is a few times $10^{-4}$ in our 2D simulations.

\item Magnetized disk winds persist despite of the VSI turbulence. The bulk wind properties remain similar to the cases without the VSI, with small variability at the level of $\sim10-30\%$. The midplane current sheet becomes corrugated as a result of the VSI, and poloidal magnetic field shows bunching in the surface layer.

\item The growth rate of the VSI depends on disk magnetization. Stronger fields lead to slower growths, and the dominant behavior transitions from corrugation to low-amplitude breathing modes. 

\item Level of VSI turbulence weakens as gas becomes better coupled to magnetic fields as long as the MRI remains suppressed. We speculate weakest turbulence is achieved at some intermediate values of $Am$.  

\item The conclusions above apply to locally isothermal disks. Finite thermal relaxation timescale weakens the VSI in a way similar to the pure hydrodynamic case, and eventually suppresses the VSI for $\tau\gtrsim0.1\mathrm{P_{orb}}$ for our fiducial disk parameters.
\end{itemize}

\subsection{Discussion}
The development of MHD winds and of hydrodynamic instabilities have been mostly studied independently in literature. This work shows for the first time that they can coexist. A profound implication of this result is that unlike the conventional thinking turbulence serves to both transport angular momentum transport and provide particle stirring, the picture now becomes that angular momentum transport is mainly driven by magnetized disk winds, whereas particle stirring is due to turbulence. This means that observational constraints on disk turbulence from either direct measurement (e.g., \citealp{flaherty_etal17,flaherty_etal18,teague_etal18}) or indirect inference from the thickness of the dust layer (e.g., HL Tau; \citealp{pinte_etal16}) do not necessarily tell the mechanism of angular momentum transport, as has already been speculated in the case of HL Tau \citep{hasegawa_etal17}.

More broadly, magnetic field tend to play a destabilizing role in the ideal MHD limit due to the development of the MRI turbulence. This has been shown by \citet{lk11} in the case of subcritical baroclinic instability. For the VSI, we have also seen that it tends to be weaker as ambipolar diffusion weakens, and it is likely eventually taken over by the MRI. Therefore, we may speculate that with a magnetically-driven disk wind launched from disk surface, the largely magnetically-inactive PPD offers a relatively clean environment for hydrodynamic instabilities to develop, and besides the VSI, it is worth investigating the case with other hydrodynamic instabilities.

As the first numerical study incorporating magnetism to investigate the vertical shear instability, we have adopted some simplifications aiming to make the simulations as clean as possible. Meanwhile, our simulations are subject to several limitations both numerically and physically that can be improved in future studies. Numerically, first we employ Ohmic resistivity in all simulations at the midplane to stabilize the current sheet (Appendix \ref{app:current}). Properly following the long-term behaviour of the midplane current sheet requires 3D simulations with reasonable resolutions. Second, our 2D simulations inevitably prevent the development of vortices, which have been known to be prevalent in 3D. Third, the inner boundary becomes less stable toward longer time, which is likely related to the secular outward transport (loss) of poloidal magnetic flux, restricting our simulations to relatively short time.

Physically, we have used prescribed vertical profiles of temperature and ambipolar diffusivity, whereas more realistic calculations should properly account for ionization chemistry (e.g., \citealp{bai17}) and radiative processes in the disk (e.g., \citealp{sk16}). Moreover, wind properties are sensitive to the physics of the wind launching region, which we have treated using a simple transition calling for further improvement (e.g., \citealp{wang_etal19}). Lastly, we have neglected the Hall effect, which is also expected to play an important role in a broad range of radii, whose behavior depends on the polarity of the net poloidal field with respect to disk rotation (e.g., \citealp{w07}). For the aligned case, prior global 2D \citep{bai17,bethune_etal17} and local 2D and 3D \citep{lesur_etal14,bai14,bai15,simon_etal15} simulations indicate that the disk is magnetically laminar, with horizontal field substantially amplified by the Hall-shear instability (HSI, \citealp{kunz08}). We might expect the VSI to coexist on top of the wind as found in this work, though it remains to be seen the outcome as the system configuration can be asymmetric about the midplane due to the HSI \citep{bai17}. For the anti-aligned case, the disk might undergo bursty turbulent motions \citep{simon_etal15}, for which further investigation needs to be pursued in global 3D simulations to understand its interplay with the VSI.

Future works are planned to extend the simulations to 3D and to incorporate more physics, which will likely yield more complete and realistic picture of gas dynamics in the outer region of PPDs.

We thank Min-Kai Lin for useful discussions and an anonymous referee for thoughtful comments following a thorough reading of the manuscript. CC acknowledges support from the Natural Science Foundation of China (grants 11573051, 11633006, 11650110427, and 11661161012). XNB acknowledges support from the Youth Thousand Talent program. This work makes use of the High Performance Computing Resource in the Core Facility for Advanced Research Computing at Shanghai Astronomical Observatory.

\appendix
\section{Resolution Study} \label{app:res}

\begin{figure*}[ht]
\epsscale{1.15}  
\plotone{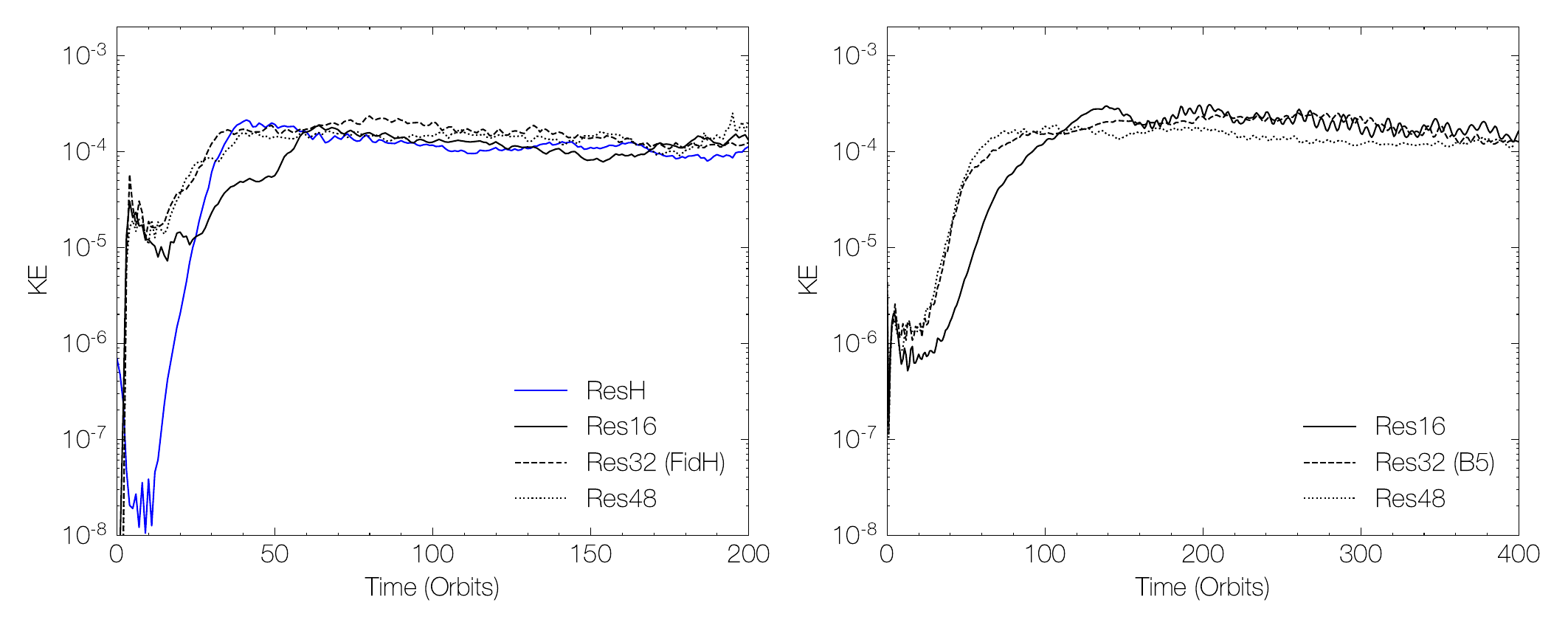}
\caption{Perturbed kinetic energies for resolution study with 16 (solid), 32 (dashes), and 48 (dotted) cells per $H_\mathrm{d}$ at the midplane. Left panel: time evolution of kinetic energies of hydrodynamic simulations.  Right panel: time evolution of perturbed kinetic energies of MHD simulations with magnetization $\beta_0=10^5$.}
\label{fig:appres}
\end{figure*}

Firstly, we compare hydrodynamic simulations via four levels of resolution. Model ResH has a domain and resolution similar to those of N13. The limited spatial domain spans over $r\in[1,4]$ and $\theta\in[-5H_\mathrm{d}/r, 5H_\mathrm{d}/r]$, and the resolution achieves 96 cells per $H_\mathrm{d}$ in $r$ and 108 cells per $H_\mathrm{d}$ in $\theta$ with uniform grid spacing. We do not apply a transition from the disk zone to the wind zone for model ResH. The other three possess large domain size the same as simulations presented in Table \ref{table:runs}, reaching 16, 32, and 48 cells per $H_\mathrm{d}$ at the midplane, respectively. The direct comparison of the evolution of kinetic energies in a box of $r\in[2,4]$ and $\theta\in[-3H_\mathrm{d}/r,3H_\mathrm{d}/r]$ are shown in the left panel of Figure \ref{fig:appres}. In the early stages, relaxation from the inner boundary dominates the fluctuations in model Res16, Res32, and Res48. After $\sim15\rm P_0$, fluctuations by the VSI take over, and the kinetic energies converge for all four runs after $\sim50\rm P_0$.

We also perform resolution study for MHD simulations but with weak disk magnetization ($\beta_0=10^5$). Three resolutions are tested as in hydrodynamical simulations. The right panel of Figure \ref{fig:appres} indicates that the three simulations have their kinetic energies converge after $100\rm P_0$. As experimented, a resolution of 48 cells per $H_\mathrm{d}$ can be time-consuming, especially when field strengths are strong. Overall, we conclude that a moderate resolution of 32 cells per $H_\mathrm{d}$ can be well suited for the purpose of this study. 

\section{Corrugation of the midplane current sheet} \label{app:current}

In our simulations, the current sheet emerges at the midplane due to the orbital shear generated toroidal magnetic fields. The current sheet dissipates toroidal magnetic field via reconnection. This current sheet is observed to be unstable and leads to fluctuations to the gas motion and magnetic field above/below the midplane. The unstable current sheet can further adversely affect the development of the VSI. As the first numerical experiment to explore the interplay between the VSI and laminar MHD winds, we implement Ohmic resistivity in the midplane to stabilize the current sheet, allowing for the VSI to develop on top of a clean background.

\begin{figure*}[ht]
\epsscale{1.2}  
\plotone{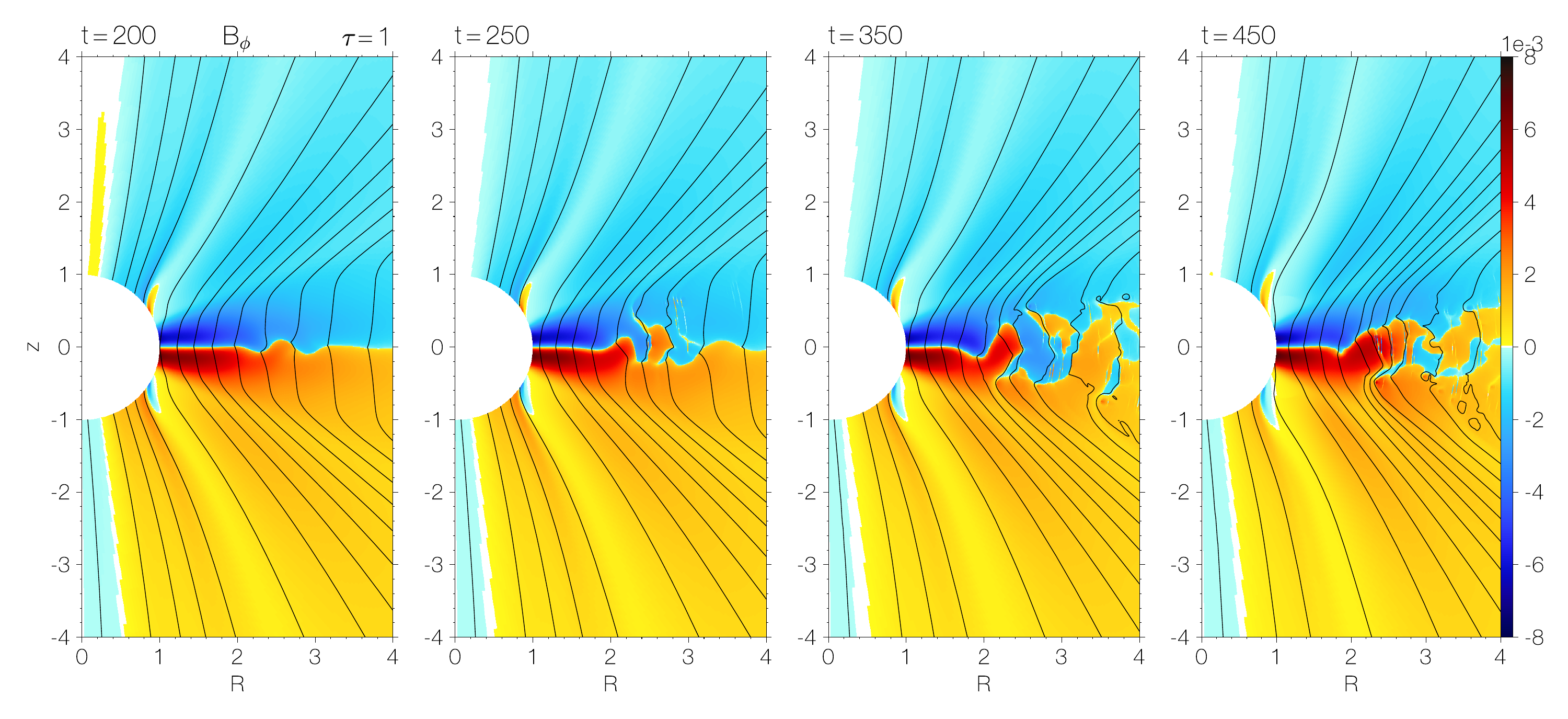}
\caption{Snapshots of toroidal magnetic field contour maps taken at $t = 200, 250, 350$, and $450\rm P_0$ of model with $\tau=1$, $Am=0.5$ and without midplane resistivity. Solid black curves show the evenly spaced contour lines of poloidal magnetic fluxes.}
\label{fig:bphiapp}
\end{figure*}

To track the behavior of the current sheet, we conduct test runs without the VSI to explore how this current sheet will evolve and affect its vicinity. Figure \ref{fig:bphiapp} shows snapshots of toroidal magnetic fields overlaid with equally spaced contour lines of poloidal magnetic flux with $\tau=1$ and $Am=0.5$. Now the resistivity at the midplane is removed while it is still retained in the buffer zone at $r<1.5$. Noticeable corrugations of current sheet in vertical direction is detected after $\sim 200\rm P_0$. As seen in later times, the level of corrugations exacerbate and break up the current sheet for $R>2$. We are currently working on 3D simulations to further investigate this issue.

\end{document}